\documentclass[twocolumns]{aastex62}

\usepackage{color}
\usepackage{graphicx}
\usepackage[normalem]{ulem}

\newcommand{\bzcat}{Roma-BZCAT}
\newcommand{\fer}{{\it Fermi}}
\newcommand{\wse}{{\it WISE}}

\graphicspath{{./}{figures/}}

\received{}
\revised{}
\accepted{}

\shorttitle{Blazar candidates in the infrared sky}
\shortauthors{D'Abrusco et al.}
\begin{document}

\title{Two new catalogs of blazar candidates in the \wse\ infrared sky}

\correspondingauthor{Raffaele D'Abrusco}
\email{rdabrusc@cfa.harvard.edu}

\author[0000-0003-3073-0605]{Raffaele D'Abrusco}
\affil{Center for Astrophysics \textbar\ Harvard \& Smithsonian, 60 Garden Street, Cambridge, MA 20138, USA}

\author{Nuria \'Alvarez Crespo}
\affiliation{Dipartimento di Fisica, Universit\`{a} degli Studi di Torino, via Pietro Giuria 1, I-10125 Torino, 
Italy}
\affiliation{European Space Agency (ESA), European Space Astronomy Centre (ESAC), E-28691 Villanueva de la 
Ca\~{n}ada, Madrid, Spain}

\author{Francesco Massaro}
\affiliation{Dipartimento di Fisica, Universit\`{a} degli Studi di Torino, via Pietro Giuria 1, I-10125 Torino, 
Italy}
\affiliation{Istituto Nazionale di Fisica Nucleare, Sezione di Torino, via Pietro Giuria 1, I-10125 Torino, Italy}
\affiliation{INAF - Osservatorio Astrofisico di Torino, via Osservatorio 20, I-10025 Pino Torinese, Italy}

\author{Riccardo Campana}
\affiliation{INAF - Osservatorio di Astrofisica e Scienza dello Spazio, via Gobetti 93/3, I-40129, Bologna, Italy}

\author{Vahram Chavushyan}
\affiliation{Instituto Nacional de Astrof\'isica, \'Optica y Electr\'onica, Apartado Postal 51 y 216, 72000
Puebla, M\'exico}

\author{Marco Landoni}
\affiliation{INAF - Istituto Nazionale di Astrofisica, via Emilio Bianchi 46, I-23807 Merate (LC), Italy}

\author{Fabio La Franca}
\affiliation{Dipartimento di Matematica e Fisica, Universit\`a Roma Tre, via della Vasca Navale 84,
I-00146, Roma, Italy}

\author{Nicola Masetti}
\affiliation{INAF - Osservatorio di Astrofisica e Scienza dello Spazio, via Gobetti 93/3, I-40129, Bologna, Italy}
\affiliation{Departamento de Ciencias F\'isicas, Universidad Andr\'es Bello, Fern\'andez Concha 700, Las Condes, 
Santiago, Chile}

\author{Dan Milisavljevic}
\affiliation{Department of Physics and Astronomy, Purdue University, 525 Northwestern Avenue, West Lafayette, 
IN, 47907, USA}

\author{Alessandro Paggi}
\affiliation{Dipartimento di Fisica, Universit\`{a} degli Studi di Torino, via Pietro Giuria 1, I-10125 Torino, Italy}
\affiliation{Istituto Nazionale di Fisica Nucleare, Sezione di Torino, via Pietro Giuria 1, I-10125 Torino, Italy}
\affiliation{INAF - Osservatorio Astrofisico di Torino, via Osservatorio 20, I-10025 Pino Torinese, Italy}

\author{Federica Ricci}
\affiliation{Instituto de Astrof\'isica and Centro de Astroingenier\'ia, Facultad de 
F\'isica, Pontificia Universidad Católica de Chile, Casilla 306, Santiago 22, Chile}

\author{Howard A. Smith}
\affil{Center for Astrophysics \textbar\ Harvard \& Smithsonian, 60 Garden Street, Cambridge, MA 20138, USA}



\begin{abstract}

We present two catalogs of radio-loud candidate blazars whose \wse\ mid-infrared 
colors are selected to be consistent with the colors of confirmed $\gamma$-ray emitting 
blazars. The first catalog is the improved and expanded release of the WIBRaLS catalog presented 
by~\cite{dabrusco14}: it includes sources detected in all four \wse\ 
filters, spatially cross-matched with radio source in one of three radio surveys
and radio-loud based on their $q_{22}$ spectral parameter. WIBRaLS2 includes 9541 
sources classified as BL Lacs, FSRQs or mixed candidates based on their \wse\ colors. 
The second catalog, called KDEBLLACS, 
based on a new selection technique, contains 5579 candidate BL Lacs extracted 
from the population of \wse\ sources detected in the first three \wse\ passbands 
([3.4], [4.6] and [12]) only, whose mid-infrared colors are similar to those of 
confirmed, $\gamma$-ray BL Lacs. 
KDBLLACS members area also required to have a radio counterpart and be 
radio-loud based on the parameter $q_{12}$, defined similarly to $q_{22}$ used 
for the WIBRaLS2. We describe the properties 
of these catalogs and compare them with the largest samples of confirmed 
and candidate blazars in the literature. We crossmatch the two 
new catalogs with the most recent catalogs of $\gamma$-ray sources 
detected by \fer\ LAT instrument. Since spectroscopic observations of candidate 
blazars from the first WIBRaLS catalog within the uncertainty
regions of $\gamma$-ray unassociated sources confirmed that $\sim\!90\%$ of 
these candidates are blazars, we anticipate that these new catalogs will play again 
an important role in the identification of the $\gamma$-ray sky.

\end{abstract}



\section{Introduction}
\label{sec:intro}

Blazars represent one of the most extreme classes of active galactic nuclei (AGNs). These radio-loud sources 
are characterized by flat radio spectra even at low radio frequencies (i.e., 
below $\sim$1\,GHz)~\citep{massaro13a,massaro13b,nori14,giroletti16}, superluminal 
motions~\citep[see e.g.,][and references therein]{vermeulen94,lister2005,lister2009}, 
peculiar infrared colors~\citep[][hereinafter Paper I]{massaro11,dabrusco14}, high optical 
polarization~\citep[see e.g.,][]{agudo14,pavlidou2014,angelakis2016,hovatta2016} and, not least, 
rapid and irregular variability~\citep[see e.g.,][]{homan02} at all frequencies with 
uncorrelated  
amplitudes and 
different time scales ranging between minutes to weeks~\citep{homan02}. Blazars can 
reach bolometric luminosities up 
to $10^{49}$ erg s$^{-1}$ during $\gamma$-ray flaring states~\citep{orienti14}. 

The emission from 
blazars, according to the unification scenario of radio-loud AGNs, arises from a relativistic
jet, closely aligned along the line of sight, that, in some cases, can outshine the host 
galaxy and the other AGN emission components~\citep[see e.g.,][]{blandford78,urry95}.

A close connection between the $\gamma$-ray and the infrared (IR) properties of blazars has 
recently emerged from the investigation of the all-sky surveys carried out with Wide-Field Infrared Survey 
Explorer (\wse)~\citep{wright10} and \fer\ satellites~\citep{dabrusco12,massaro16}. The observed
correlations across more than 10 
decades of frequency has been proven to be an extremely powerful tool to identify new $\gamma$-ray emitting 
blazars among \fer\ sources with unknown or uncertain lower-energy 
counterparts~\citep{massaro12a,dabrusco13}.

With a sky density of $\sim\! 0.1$ sources per square degree, blazars dominate the $\gamma$-ray 
sky in the MeV-TeV energy range and represent $\sim\!40\%$ of the sources in the Third \fer\ -Large Area 
Telescope (LAT) Source Catalog~\citep[3FGL,][]{acero15,massaro15c}. Blazars are mainly divided into two 
sub-classes based on their optical spectra: (i) BL Lac objects characterized by featureless optical 
spectra or 
showing emission and/or absorption lines of equivalent widths $EW<$ 5\AA\ ~\citep{stickel91,falomo14}, and (ii)
Flat Spectrum Radio Quasars (FSRQs) with a typical quasar-like optical spectrum. Hereinafter, we adopt the 
nomenclature used in the \bzcat\ catalog~\citep{massaro09,massaro15a}, where BL Lacs and FSRQs are 
referred to as BZBs and BZQs, respectively.

The Spectral Energy Distributions (SEDs) of blazars are characterized by two broad components, a  
low-energy one peaking between the IR and X-ray bands and a high energy one whose emission ranges between
the X-rays and the $\gamma$-rays. The former component is interpreted as synchrotron 
radiation from relativistic particles accelerated in a jet while the component at higher energies, according to leptonic models, is due to inverse Compton emission with seed photons that can 
have different origins~\citep[for more details see e.g.,][]{bottcher07,bottcher12}, while hadronic 
models invoke the synchrotron emission by protons or secondary particles produced in proton-photon 
interactions~\citep[][and references therein]{dermer1993,mucke2001,murase2012}.

The BL Lac 
population is also divided, according to their SEDs, in ``Low-frequency peaked BL Lac objects'' 
(LBLs) when the peak of the first 
component lie in the IR-to-optical energy range and ``High-frequency peaked BL Lac objects'' (HBLs) when 
the synchrotron peak falls in the UV-to-X-rays energy range~\citep{padovani96}. Another, more recent 
classification distinguishes blazars as low-synchrotron peaked (LSP), intermediate-synchrotron peaked (ISP) 
or high-synchrotron peaked (HSP) based on the peak frequency $\nu_{S,\mathrm{peak}}$ of the synchrotron 
component of their SEDs~\citep{abdo10,ackermann15}. Here we will adopt the LBLs/HBLs sub-classification for 
the BZBs as it does not strictly depend on the exact location of the peak frequency.

The discovery of the peculiar \wse\ IR colors of blazars has been used to search for blazar-like sources within 
the positional uncertainty regions of the unidentified/unassociated $\gamma$-ray sources (UGSs) that could 
be their potential counterparts~\citep[see e.g.,][]{massaro15b}. Several procedures based on \wse\ data 
have been developed 
to investigate the nature of the UGSs listed in all \fer\ source catalogs as well as to verify the nature of blazars
candidates of uncertain type~\citep[BCUs;][]{massaro12b,cowperthwaite13,alvarez16a}. However, all these methods 
require spectroscopic confirmation of the natures of the blazar candidates selected.

One of the largest sources of candidate blazars used to identify unassociated $\gamma$-ray sources 
observed by the \fer\ 
LAT has been the catalog of \wse\ Blazar-Like Radio-Loud Sources~(WIBRaLS; Paper I). 
This catalog contains \wse\ sources detected in all four \wse\ bands, whose 
mid-IR colors are similar to those of confirmed \fer\ blazars. WIBRaLS sources were also required to a) have a radio 
counterpart from one of three major surveys, namely the National Radio Astronomy Observatories Very Large Array VLA Sky 
Survey~\citep[NVSS,][]{condon98} the VLA Faint Images of the Radio Sky at 
Twenty-cm Survey~\citep[FIRST,][]{white97,helfand15}
and the Sydney University Molonglo Sky Survey Source Catalog~\citep[SUMSS,][]{mauch03}, and b) to be radio-loud, 
that is to have an observed ratio between radio and 22 $\mu$m mid-infrared flux densities $>3$ (Paper I).

Since the publication of the WIBRaLS catalog, extensive optical spectroscopic campaigns whose goal 
is to verify the
nature of WIBRaLS1 blazar candidates that can be spatially associated to $\gamma$-ray sources observed 
by \fer\ have been carried out~\citep[see e.g.,][]{massaro14,paggi14,lamura2015,landoni15,massaro15spectra,ricci15,ajello2017b,alvarez16a,alvarez16b,alvarez16c,pena17,paiano2017a,paiano2017b,landoni2018,marchesi2018,paiano2019,marchesini2019}.
The total number of WIBRaLS candidate blazars which have been spectroscopically followed up and reported 
upon
in either one of the papers listed above is, to date, 159 split in 126 candidate BZBs, 16 candidate BZQs and 17 
candidates with no spectral classification available (Mixed). The analysis of the optical spectra 
confirmed that 
$\sim\!93\%$ of candidate BZBs have featureless optical spectra typical of BL Lacs and $\sim\!52\%$ of the 
candidate BZQs show FSRQ spectra. Only $\sim\!3\%$, $\sim\!12\%$ and $\sim\!11\%$ of the spectra of observed 
candidate BZBs, BZQs
or Mixed blazars cannot be classified as belonging to blazars, yielding a weighted average efficiency
of the selection $\sim\!95\%$. 
Checks of spectra already published in the literature for 28 additional candidate blazars
from WIBRaLS1 confirmed a BL Lacs or FSRQs nature for 27 of them.

Since the publication of the first release of the WIBRaLS catalog~(Paper I), new versions of 
some of the main datasets used to 
define its selection method have been released:

\begin{itemize}
    \item The ROMA BZCat, which contains the list of {\it bona fide}, spectroscopically confirmed blazars 
    used to define the {\it locus} occupied by $\gamma$-ray emitting blazars in the \wse\ color space, 
    has reached its 5th release~\citep{massaro09}. This release contains $\sim$ 3600 sources, vs $\sim$ 
    3050 sources in the version used by~\cite{dabrusco14}. 
    \item The LAT 3-year Point Source Catalog (3FGL)~\citep{acero15} of 
    $\gamma$-ray sources detected by \fer\, containing $\sim$ 3000 sources, has also become available. 
    This recent update to the catalog of \fer\ LAT sources is more than twice as large as the 2-year 
    2FGL~\citep{nolan12} catalog ($\sim$ 2250 members) used to extract the first version of the 
    WIBRaLS catalog.
\end{itemize}

These two new datasets yield jointly a larger sample of confirmed 
$\gamma$-ray blazars that can be used to characterize more accurately their \wse\ mid-IR properties and
hence to identify more effectively blazars that may be detected in the $\gamma$-ray 
energy range. In this paper we describe a 
new release of the WIBRaLS catalog that, by taking advantage 
of these most recent data available, maximizes the legacy value of \wse\ observations for the investigation 
of blazars.

The completeness of the WIBRaLS catalog is a function of the blazar spectral class, and decreases significantly 
for BL Lacs, which often are not detected in the $[22]$ $\mu$m \wse\ band. BZBs, in particular 
HBLs, have lower detection rate in the $[22]$ \wse\ band~(Paper I), since their emission 
in the mid-infrared at 22 $\mu$m may be lower than the limiting sensitivity in the fourth \wse\ band. 

For this reason, in this paper we will also present a new, complementary 
catalog of candidate BZBs selected with a novel technique that employs the \wse\ colors obtained from 
the first three \wse\ bands only and has been applied to All\wse\ sources not detected in W4.
We focus only on the BZB spectral class for two main reasons: i) the region of the IR color-color space occupied 
by BZB is less contaminated by spurious IR sources than that of BZQs, ii) during our campaign of spectroscopic 
follow-up of candidate BL Lacs selected with \wse\ colors, we found that a large fraction of UGSs and BCUs are 
classified as candidate BZBs, providing an indication that they are the most elusive counterparts of \fer\ 
sources~\citep[70.5 \% and 65.4 \%, respectively, see][]{massaro16b}.

The paper is organized as follows: in Section~\ref{sec:data} we provide a brief introduction to the \wse\ and 
radio catalogs used to extract both samples of candidate blazars. Sections~\ref{sec:wibrals} 
and~\ref{sec:wisekde} describe the selection methods of the new WIBRaLS catalog and the catalog of BL Lac 
candidates selected using two \wse\ colors only, respectively. The comparison between the two 
catalogs presented in this paper and the literature is described in 
Section~\ref{sec:catalogs}. Finally, our conclusions are summarized in Section~\ref{sec:summary}. 

We use cgs units unless otherwise stated and spectral indices, $\alpha$, are defined by flux density 
$S_\nu \propto \nu^{-\alpha}$ indicating as flat spectra those with $\alpha\!<\!0.5$. \wse\ magnitudes used 
here are in the Vega system and are not corrected for the Galactic extinction. As shown in our previous 
analyses~\citep{dabrusco13,dabrusco14}, such correction affects only the magnitude at 3.4 $\mu$m for sources 
lying at low Galactic latitudes, and it ranges between 2\% and 5\% of the magnitude, thus not affecting 
significantly our results. 

\section{Data}
\label{sec:data}

\subsection{Infrared data}
\label{sec:data_wise}

All the candidate blazars discussed in this paper were sources extracted from images produced by the \wse\ space 
telescope~\citep{wright10}. \wse\ has observed the whole sky from 2009 to 2011 in the four bands W1, W2, W3 
and W4 centered on 3.4, 4.6, 12 and 22 $\mu$m, respectively. 

The All\wse\ source catalog has superseded the previously available \wse\ All-Sky catalog; it was 
produced by combining \wse\ single exposures images from the first two years of the mission with the 
post-cryogenic phase data~\citep{mainzer11}, that observed the sky in three (W1, W2 and W3) and then only 
two (W1 and W2) bands. The final result is a two-fold improvement in the depth-of-coverage in 
the first two bands thanks to the additional observations that have increased the sensitivity of the stacked 
images, and improved photometric accuracy in all filters thanks to updated background measurements. The All\wse\
sensitivities are 54, 71, 730 and 5000 $\mu$Jy in the W1, W2, W3 and W4 passbands respectively, with 
angular resolutions of 6\arcsec.1, 6\arcsec.4, 6\arcsec.5 and 12\arcsec. Due to the \wse\ survey strategy, 
the limiting sensitivity of the catalog of sources extracted from the \wse\ images is not uniform on the sky 
(compare with Figure~8 at the Explanatory Supplement to the All\wse\ Data Release 
Products\footnote{\url{http://wise2.ipac.caltech.edu/docs/release/allwise/expsup/sec4_2.html}}). 

The All\wse\ Source Catalog contains astrometry and photometry in the IR for 747,634,026 objects; 
only 25,882,082 of these sources ($\sim$3.5\%) were seen in all the four bands, increasing up to 
99,118,890 sources ($\sim$13.3\%) 
detected in the first three bands W1, W2 and W3 only. These two catalogs represent the parent samples for the 
updated WIBRaLS (Section~\ref{sec:wibrals}) and the new catalog of BL Lacs candidates (Section~\ref{sec:wisekde}), 
respectively. 

\subsection{Radio data}
\label{sec:data_radio}

We have crossmatched the catalog of All\wse\ sources according to the procedure described in 
Section~\ref{sec:radctp} with the NVSS, the FIRST and the SUMSS radio surveys. A brief description
for each of these three radio surveys is given below.

\begin{itemize}
    \item NVSS~\citep{condon98} is a 1.4 GHz continuum survey performed using the Very Large Array (VLA) 
    and covering the entire sky north of $-40$ deg declination, i.e. 82\% of the celestial sphere with 
    a beam size of 45\arcsec\ FWHM. The result is a catalog of over 1.8 million discrete sources brighter 
    than S $\sim$ 2.5 mJy in the entire survey. 
    \item FIRST~\citep{white97,helfand15} is a project designed to produce the radio equivalent of the 
    Palomar Observatory Sky Survey over 
    10,000 square degrees of the North and South Galactic Caps using the NRAO Very Large Array (VLA). 
    The beam size varies between 5\arcsec.4 FWHM for circular beam and 6\arcsec.8 along the major axis
    for elliptical shape, as a function of the declination of the observation. The 
    survey area has been chosen to coincide with that of the Sloan Digital Sky 
    Survey~\citep[SDSS, see e.g.,][]{gunn1998} and at the m$_v$ $\sim$23 limit of SDSS, $\sim$40\,\% of the 
    optical counterparts to FIRST sources are detected. At the 1 mJy source detection threshold, there 
    are $\sim$90 sources per square degree.
    \item SUMSS~\citep{mauch03} is a radio imaging survey of the southern sky carried out with the Molonglo 
    Observatory Synthesis Telescope (MOST) operating at 843 MHz, with a beam size $\approx$ 43\arcsec\ FWHM. 
    The catalog covers approximately 3500 deg$^2$ with declination $\delta <$ 
    $-$30$^\circ$, about 43\% of the total survey area. The survey has a limiting peak brightness of 6 mJy/beam at 
    declinations $\delta \leq$ $-$50$^\circ$, and 10 mJy/beam at $\delta >$ $-$50$^\circ$. The SUMSS is therefore 
    similar in sensitivity and resolution to the NVSS, with $\sim\!$7000 sources found in the overlap region.
\end{itemize}

\section{The second release of the WIBRaLs catalog}
\label{sec:wibrals}

In this paper, a new release of the catalog of \wse\ Blazar-like RAdio-Loud Sources (WIBRaLS) has 
been built up by following an improved version of the procedure described by Paper I and including
new samples of confirmed blazars associated with \fer\ $\gamma$-ray sources. Schematically, the steps we followed 
to extract the WIBRaLS catalog are the following: 

\begin{enumerate}
	\item We select \wse\ sources detected in all four \wse\ bands, with IR colors similar to those of associated 
	\fer\ blazars (Section~\ref{sec:pc}). We call these sources ``\wse\ blazar-like sources''. 
	\item The \wse\ blazar-like sources are positionally cross-matched with sources extracted from either one
	of the radio surveys NVSS, SUMSS and FIRST (Section~\ref{sec:radctp}) and only those with a radio
	counterpart are retained.
	\item Among the WISE blazar-like sources with a radio counterpart, only radio-loud sources 
	are selected as members of the WIBRaLS catalog (Section~\ref{sec:radioloud}).
\end{enumerate}	

In the following, we provide descriptions of each step above.

\subsection{\wse\ color selection of blazar-like sources}
\label{sec:pc}

We define the blazar-like {\it locus} using the sample of confirmed {\it Fermi} blazars, based on the 5$^{th}$ version of 
the ROMA-BZCat catalog~\citep{massaro15a} and the 3FGL catalog~\citep{acero15}, associated with All\wse\ 
counterparts detected in all four \wse\ filters. This {\it locus} thus includes newly identified 4-band
sources and extends the {\it locus} presented in Paper I.

The BZCat v.5.0 contains 3561 {\it bona fide} blazars with spectroscopic confirmation. Optical spectra are 
also used to classify members of the BZCat as BZBs or BZQs, according to the total equivalent width of 
all emission and absorption features, while blazars whose properties are intermediate 
between BZB and BZQ are tagged as Uncertain (BZU) and other blazar-like objects whose emission is
mostly contaminated by light from the host galaxy are labeled as BZGs. 

All BZCat sources classified as BZU and BZG were discarded, leaving 1151 BZBs and 1909 BZQs. In order to 
spatially crossmatch the positions of these BZCat sources with All\wse\ sources detected in the four 
filters, we adopted the maximum radial distance of 3\arcsec.3 obtained by conservatively combining a nominal
uncertainty of 1\arcsec on the radio positions of BZCat sources with positional uncertainty in the \wse\ W4
passband~\citep[see][for details]{dabrusco13}. This sample was then spatially crossmatched with $\gamma$-ray 
sources included in the 
3FGL catalog~\citep{acero15}, the latest release of sources detected by the LAT instrument 
on board NASA's \fer\ spacecraft, based on the first 48 months of survey data. The crossmatch has been performed 
by taking into account the position angles and the lengths of the semi-major and semi-minor axes of the 95\% 
confidence region of each \fer\ source, available in 3FGL, and the positional uncertainty of the All\wse\ 
counterparts of the BZCat sources. The final sample used to define the \wse\ three-dimensional {\it locus} 
consists of 901 confirmed $\gamma$-ray emitting blazars, split in 497 BZBs and 404 BZQs, and is more than twice
as large as the analogous list in Paper I, which contained 447 sources.

Following Paper I, we define a model of the {\it locus} in the three-dimensional Principal Components 
(PCs) space generated by the three independent \wse\ colors W1-W2 ($[3.4]$-$[4.6]$), W2-W3 ($[4.6]$-$[12]$) and 
W3-W4 ($[12]$-$[22]$). 
The {\it locus} is modeled as a set of three coaxial cylinders whose axes lie on the direction of the first 
PC (PC1). 
The two extremal cylinders are populated by blazars of similar spectral classes, namely BZB 
and BZQ respectively, while the intermediate cylinder contains significant ($\geq\!25\%$) fractions of 
both BZBs and BZQs. The length along the PC1 axis of the two extremal cylinders is defined so that 
they contain at least 75\% of blazars classified as candidate BZBs and BZQs, respectively, while 
excluding the {\it locus} sources with PC1 value smaller than the 1\%-st percentile and larger than 
the 99\%-th percentile of the distribution of {\it locus} PC1 coordinates.
The mixed cylinder contains fractions of BZBs and BZQs each smaller than 75\%. The radii of the three 
cylinders, which 
lie in the plane generated by PC2 and PC3 and orthogonal to PC1, are defined to contain 95\% of the 
sources whose PC1 coordinates fall in each of the three cylinders. Sources located within the 
{\it locus} are selected on the base of the value of their ``score'', a quantitative measure of the distance 
of a generic source to the {\it locus} model, defined as follows.

The \wse\ colors of a \wse\ source and the associated uncertainties are 
projected into the PC three-dimensional space, where they define an ``uncertainty ellipsoid''. The position and
orientation of this ellipsoid - relative to the model of each cylinder separately - are used to calculate a numeric 
value defined between 0 and 1. This normalized distance is then weighted by the volume of the error ellipsoid, 
so that two \wse\ sources placed in the same position relative to the {\it locus} model but with different 
uncertainties are assigned different scores. A more detailed description of the score and its properties 
can be found in Paper I. 

The score is also used to classify candidate blazars compatible with the model of the {\it locus} 
according to their reliability. All \wse\ sources with non-zero score for either cylinder 
are ranked according to the decreasing compatibility with the {\it locus} model in the classes A, B, C and D, 
defined by the 90\%-th, 60\%-th, 20\%-th and 5\%-th percentiles of the {\it score} distribution for the sources 
of the {\it locus} sample. These classes are defined to facilitate quick prioritization of 
candidate blazars as targets of follow-up observation across sources of different spectral types (BZB-like,
BZQ-like or Mixed) and regardless of the specific {\it score} distributions for each spectral type.
Nonetheless, classes do not replace {\it scores} as a quantitative indicator of the degree of compatibility 
of each candidate blazar with the {\it locus} model in the \wse\ color space; all scientific 
analysis on the final list of WIBRaLS should be performed using {\it scores}.
The lower 5\% threshold corresponds to the fraction of {\it locus} sources that are 
located outside the {\it locus} model by definition. The values of the {\it score} thresholds used to define 
the three classes are reported in Table~\ref{tab:thresholds}. 

To all sources with score larger than the 5-th percentile threshold, for any one of the three cylinders, are assigned 
the corresponding type (BZB, BZQ or Mixed) and selected as \wse\ blazar-like sources. The Mixed type does 
indicate a specific spectral class, since the Mixed cylinder contains comparable fractions of both BZBs and BZQs.

The total number of \wse\ blazar-like sources selected is 526,681, split in 156,506 BZB candidates 
($\sim\!30\%$), 348,805 ($\sim\!60\%$) BZQ candidates and the remaining 21,370 ($\sim\!4\%$) sources located 
in the Mixed region. The \wse\ blazar-like sources can also be split in 7,807 Class A sources ($\sim\!1\%$), 27,986 
Class B sources ($\sim\!5\%$), 149,052 Class C sources ($\sim\!28\%$) and the remaining 341,836 
($\sim\!65\%$) belonging to the class D.

\begin{deluxetable}{cccc}
	\tablecaption{Values of the {\it score} thresholds $s_{5\%}$, $s_{20\%}$, $s_{60\%}$ and $s_{90\%}$, 
	used for the extraction of the \wse\ blazar-like sources and to define the classes
	as described in Section~\ref{sec:pc}}. These values are determined as the 5\%-th,
	20\%-th, 60\%-th and 90\%-th percentiles of the distribution of {\it scores} of the {\it locus} 
	sample, separately for BZB, Mixed and BZB cylinders in the {\it locus} model.\label{tab:thresholds}
    \tablehead{
	    & \colhead{BZB-like} 	& \colhead{Mixed}	& \colhead{BZQ-like}  
	    }				
    \startdata
	$s_{5\%}$ (D)	& 0.24	& 0.32	& 0.22	\\
	$s_{20\%}$ (C)	& 0.51	& 0.54	& 0.48	\\
	$s_{60\%}$ (B)	& 0.74	& 0.85	& 0.77	\\
	$s_{90\%}$ (A)	& 0.91	& 0.99	& 0.91 	\\	
	\enddata
\end{deluxetable}

\subsection{Radio counterparts}
\label{sec:radctp}

Following Paper I, we determined the optimal radii for the association of the \wse\ blazar-like sources 
with their potential radio counterpart in the NVSS, SUMSS and FIRST surveys. 

We adopted a modified version of the procedure illustrated by~\citet{best05} and~\citet{donoso09}. They 
computed the optimal radius for the spatial crossmatch of NVSS and FIRST radio detections with optical sources 
in SDSS by setting a threshold on the fraction of spurious associations (i.e. the contamination) obtained for different 
values of maximum crossmatch radius. In this paper, similarly at what done in Paper I, the optimal association 
radius is indeed set as 
the radial distance that provides a given fixed efficiency of the selection $e_{\mathrm{thr}}\!=\!99\%$, corresponding 
to a contamination $c_{\mathrm{thr}}\!=\!1\%$, where $c(\vartheta)\!=\!100\%\!-\!e(\vartheta)$. The efficiency or purity 
of the selection is defined as the number of sources around real radio positions $n_{\mathrm{real}}(\vartheta)$, reduced 
by the number of sources around mock radio positions $n_{\mathrm{mock}}(\vartheta)$ and divided by the number of 
``real'' matches.

We estimated $n_{\mathrm{real}}(\vartheta)$ by counting the number of \wse\ sources detected in all four bands within 
circular regions of radius $\vartheta$ between 0\arcsec\ and 60\arcsec\, centered on a sample of $5\!\cdot\!10^{4}$ sources 
randomly extracted from each of the three radio surveys. The value of $n_{\mathrm{mock}}(\vartheta)$ was calculated by 
averaging over one hundred mock realizations of the coordinates of each real radio source, generated by moving the 
real position in a random direction and by a random radial distance in the $[60, 120]$\arcsec\ range.

The maximum cross-match radii evaluated for the NVSS and SUMSS surveys are $\vartheta_{\mathrm{NVSS}}\!=\!10\arcsec.4$, 
$\vartheta_{\mathrm{SUMSS}}\!=\!7\arcsec.4$ respectively, identical to those determined for the 1$^{st}$ release of the 
WIBRaLS catalog. This approach does not work for the FIRST survey because of the very high density 
of FIRST sources. For this reason, we assume as optimal search radius for FIRST the value previously determined 
$\vartheta_{\mathrm{FIRST}}\!=\!3\arcsec.4$~(Paper I), that was obtained by combining conservative estimates 
of the positional uncertainties of FIRST and All\wse\ sources. 

The number of \wse\ blazar-like sources associated with at least one radio counterpart in either of the three surveys 
within the maximum radial distances discussed above is 32630, split in 18903 with a NVSS counterpart, 1040 with 
a FIRST counterpart and 3323 with a SUMSS counterpart. In order to exclude \wse\ sources associated to distinct 
radio sources associated to the emission of lobes of the same radio galaxy, we searched and removed all duplicate 
radio sources whose positions would fall within 6\arcmin from each other. This radius matches the typical positional 
uncertainty of $\gamma$-ray sources detected by LAT in the 3FGL~\citep{acero15}. In these cases, the All\wse\ source 
with the largest score among all the duplicates was retained. After this step, the number of \wse\ blazar-like sources
with a radio counterpart in the NVSS, FIRST and SUMSS catalogs is 18693, 10227 and 3287 respectively, for a total of 
32207 sources. This sample includes 5547 sources with a counterpart 
from both the NVSS and FIRST surveys and 709 sources with one counterpart from both NVSS and one SUMSS. 
The number of unique \wse\ blazar-like sources associated to a radio counterpart (i.e., the previous sample 
after removing sources with radio counterpart listed in two catalogs) is 25951, with $\sim\!72\%$ (18693) has a NVSS 
counterpart, $\sim\!18\%$ (4687) has a FIRST counterpart and the remaining $\sim\!10\%$ (2578) are associated 
to a FIRST source. Moreover, 7101 sources ($\sim\!27\%$) of this list are classified as candidate BZBs, 2186 ($\sim\!8\%$) 
as Mixed candidates and the remaining 1664 ($\sim\!64\%$) as candidate BZQs~\footnote{The selection 
of candidate BZQs using solely 
their \wse\ photometry can be contaminated by normal quasars as they share the same region of the \wse\ colors
space~\citep{wright10}. Also, the existence of radio counterparts to \wse\ -selected BZQ candidates alone, 
lacking radio spectral characterization, does not remove the degeneracy~\citep{stern2005,stern2012}}, 
with 1366 class A ($\sim\!5\%$), 3302 class B ($\sim\!13\%$), 9265 class C ($\sim\!36\%$) and the remaining 
12018 ($\sim\!47\%$) as class D sources.

\subsection{Radio-loudness selection}
\label{sec:radioloud}

Blazars are radio-loud AGNs but not every radio-loud AGN is a blazar. Blazars are generally hosted in 
elliptical galaxies, whose emission at radio frequencies 
is dominated by synchrotron emission from particles accelerated in 
the AGN relativistic jet pointed along the line of sight. In order to distinguish blazars from other radio 
sources, when lacking radio spectral information, we adopted the approach described in what 
follows.~\citet{padovani11} as well as~\citet{bonzini13}
suggested that AGN-powered radio-sources can be 
identified using $q_{24}$, a modified definition of the so-called $q$ parameter~\citep{helou85}, i.e. the 
logarithm of the ratio of far-IR to radio flux density, to overcome the dearth of accurate flux density measurements 
at far-IR frequency. The $q_{24}$ parameter is defined as:

\begin{equation}
	q_{24}\!=\!\log{(S_{24\mu m}/S_{1.4\mathrm{GHz}})}
\end{equation} 

\noindent where $S_{24\mu m}$ is the observed flux density at 24 $\mu$m and $S_{1.4GHz}$ is the 
flux density measured at 1.4 GHz. Following~\citet{dabrusco14}, we adopted a similar criterion to 
select ``radio-loud sources'' among 
the \wse\ blazar-like sources with a radio-counterpart by using the parameter $q_{22}$ defined as:

\begin{equation}
	q_{22}\!=\!\log{(S_{22\mu m}/S_{\mathrm{radio}})}
	\label{eq:q22}
\end{equation} 

\noindent where the flux density in the 24 $\mu$m band of the Multi-band Imaging Photometer used on 
{\it Spitzer} (MIPS) is replaced by the \wse\ W4 ($[22]$ $\mu$m) band. This approach is possible 
thanks to the similarities of the two wavebands~(see Paper I for additional details). 
For the $S_{\mathrm{radio}}$ in Equation~\ref{eq:q22} we used the radio flux density at 1.4 GHz for sources with a 
NVSS or FIRST radio counterparts. Given the lack of flux measurement at 1.4 GHz for SUMSS sources, for 
those sources we used flux densities at 843 MHz. 

Due to the remarkable flatness of radio spectra of blazars~\citep[see e.g.,][]{healey07,massaro13f}, 
replacing the flux density at 1.4 GHz with the same quantity at 843 MHz produces a
small effect of the value of the $q_{22}$ parameter, whose size was
estimated using the sample of \wse\ blazar-like 
sources with a radio 
counterpart detected in both NVSS and SUMSS (508 sources). The distribution of the difference between 
the values of $q_{22}(1.4~\mathrm{GHz})$ and $q_{22}(843~\mathrm{MHz})$ is fairly constant across the 
interval of $q_{22}(1.4~\mathrm{GHz})$ covered by our sample, with $\Delta q_{22}\!\approx\!-0.07$,
confirming a fairly flat radio spectrum for the sources in this sample. Based on 
this finding, for the \wse\ blazar-like sources associated with SUMSS counterparts only, 
we have used the corrected $q_{22}$ 
value defined as $q^{'}_{22}(843~\mathrm{MHz})\!=\!q_{22}(843~\mathrm{MHz})\!+\!\Delta q_{22}$. 

 \begin{figure}[ht] 
    \centering
	\includegraphics[scale=0.5]{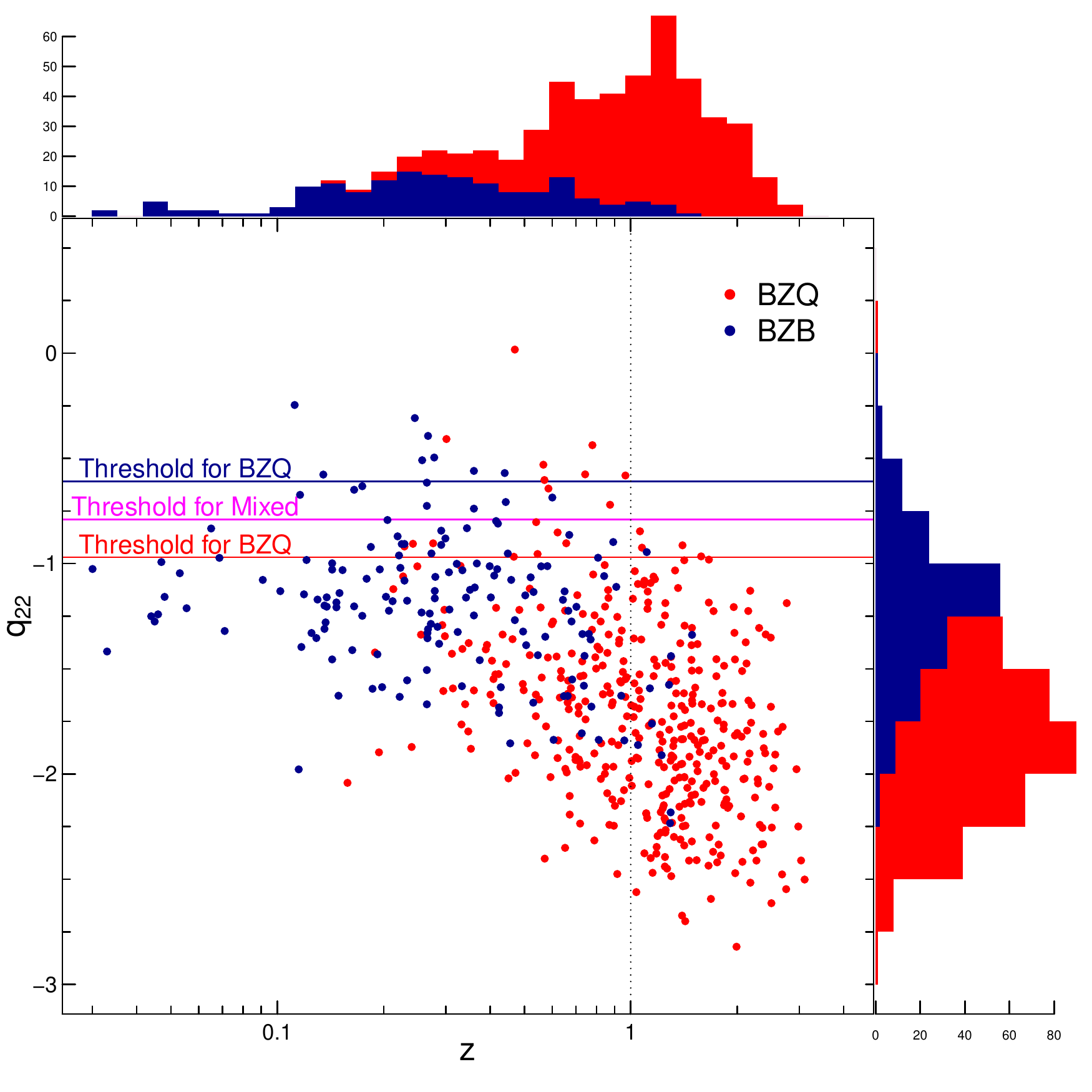}
	\caption{Scatterplot of the $q_{22}$ values for the confirmed $\gamma$-ray emitting
	blazars in the {\it locus} sample as a function of their redshifts reported in the 
	BZCat~\citep{massaro15a}. 
	The red, magenta and blue horizontal lines show the values of the $q_{22}$ thresholds
	used to select WIBRaLS sources among the \wse\ blazar-like sources with a radio counterparts
	for \wse\ -based classes of candidate BZQs, Mixed and BZBs (see Section~\ref{sec:radioloud}).}
    \label{fig:q22_locus}
\end{figure}

Both the $q_{24}$ and $q_{22}$ parameters show a dependence on the redshift of the source, as their values 
decrease for larger redshifts~\citep{bonzini13,dabrusco14}. Figure~\ref{fig:q22_locus} shows the distribution 
of $q_{22}$ values for the confirmed $\gamma$-ray emitting blazars in the {\it locus} sample with reliable 
redshift measurements, as a function of the redshift and color-coded according to their spectral classification 
from the BZCat. {\it Locus} sources (mostly BZBs) with uncertain or unknown redshifts were not used, reducing the 
number of {\it locus} members to 563 (split in 159 BZBs and 404 BZQs). In Figure~\ref{fig:q22_locus}, the same 
trend reported by~\citet{bonzini13} and Paper I is observed. If redshift estimates were available for 
all \wse\ blazar-like sources associated to a radio counterpart, we could have determined different $q_{22}$ 
thresholds in different redshift bins. Since redshifts are not available, we used fixed values of $q_{22}$. 

In this paper, we improve over the approach used in~Paper I by determining different $q_{22}$ thresholds 
for \wse\ blazar-like sources with radio counterpart classified as candidate BZQs, BZBs and Mixed. The thresholds 
for candidate 
BZBs and BZQs, calculated as the 95\% of the $q_{22}$ distribution of all blazars in the {\it locus} sample classified 
as BZB and BZQ, are $q_{22}^{\mathrm{BZB}}\!\leq\!-0.61$ and $q_{22}^{\mathrm{BZB}}\!\leq\!-0.97$, 
respectively. The threshold for Mixed sources is set to the mean value of the thresholds for candidate BZQs and BZBs,
$q_{22}^{(\mathrm{Mixed})}\!\leq\!-0.79$. These thresholds are shown as horizontal lines in Figure~\ref{fig:q22_locus}.

We evaluated the effect of the unknown underlying redshift distribution of the \wse\ blazar-like sources 
with radio counterpart on our selection based on fixed $q_{22}$ thresholds. We followed the same strategy used 
in Paper I. We computed the $q_{22}$ for all sources in the {\it locus} sample after varying their 
observed redshifts over an equally spaced grid covering the $[0, 4]$ range with bins of 0.05 width, which includes 
the interval $[0, \sim\!1.25]$ covered by the observed redshifts. We assumed power-law spectral energy distribution 
with slope constrained by the observed flux densities at 22 $\mu$m and at 1.4 GHz. The fractions of candidate 
BZQs and BZBs that satisfy the $q_{22}$ conditions based on the observed redshift distribution are 
$\sim\!91\%$ and $\sim\!93\%$, respectively. These fractions are slightly lower than the $\sim\!94\%$ fraction of 
sources recovered with the $q_{22}\!\leq\!-0.5$ threshold used in~Paper I.

\begin{deluxetable}{lcccc}
	\tablecaption{Members of the WIBRaLS catalog, split according to their 
	\wse\ spectral types and classes, and provenance of the radio counterpart.\label{tab:WIBRaLS}}
	\tablehead{
		& 	\colhead{BZB-like}	& \colhead{Mixed}& \colhead{BZQ-like}		& 	 \colhead{Total}	\\
	    }
	\startdata
	Class A	&	55	    & 5		&61			& 	121	    \\
	Class B	&	273	    &124	&317		& 	714	    \\
	Class C	&	1086	&579	&1746		& 	3411	\\
	Class D	&	2330	&0		&2965		&	5295	\\
			&		    &		&			&		    \\
	NVSS	&	3024	& 557	&4093 		& 	7664	\\
	FIRST	& 	36	    &7		&21	 		& 	64	    \\
	SUMSS	& 	694	    & 144	&975 		& 	1813	\\		
	Total 	& 	3744	&	708	& 5089		& 	9541	\\
	\enddata
\end{deluxetable} 
 
The total number of \wse\ blazar-like sources with a radio counterpart that satisfies the radio-loudness 
criteria based on the $q_{22}$ parameter and, thus, belong to the second release of the WIBRaLS catalog 
is 9541. The break-down of the catalog according to \wse\ spectral type, class and provenance of the 
radio counterpart is given in Table~\ref{tab:WIBRaLS}, while the basic parameters for a subset of sources 
in the catalog are displayed in Table~\ref{tab:cat_wibrals}. The number of sources of the {\it locus}
sample of confirmed $\gamma$-ray emitting blazars used to determine the WIBRaLS selection, that are found 
in the final WIBRaLS2 is 666, $\sim 26\%$ less than the original size of the {\it locus}
sample (901 sources). The exclusion of these 235 {\it locus} sources is the result of a) the definition of the 
\wse\ model (Section~\ref{sec:pc}), that excludes 2\% of the sources based on their location along the PC1 axis, 
and 5\% of each spectral class because the radii of the three cylinders are defined to contain the 95\% of associated
sources, and b) the definition of the $q_{22}$ thresholds (Section~\ref{sec:radioloud}), which remove 5\% of the 
remaining {\it locus} sources for each spectral class, by definition. 

\begin{deluxetable*}{lrrrrrrrrrrlrr}
   \tabletypesize{\tiny}
 	\tablecaption{Sample of rows of the catalog of WIBRaLS sources.\label{tab:cat_wibrals}}
    \tablehead{
	   \colhead{All\wse\ name\tablenotemark{a}} & \colhead{R.A.\tablenotemark{b}} & 
 	    \colhead{Dec.\tablenotemark{c}} & 
 	    \colhead{W1-W2\tablenotemark{d}} & \colhead{W2-W3\tablenotemark{e}} & 
 	    \colhead{W3-W4\tablenotemark{f}} & \colhead{$s_{\mathrm{BZB}}$\tablenotemark{g}}  & 
 	    \colhead{$s_{\mathrm{MIX}}$\tablenotemark{h}} & 
 	    \colhead{$s_{\mathrm{BZQ}}$\tablenotemark{i}} & \colhead{Class\tablenotemark{j}}  & 
 	    \colhead{Type\tablenotemark{k}} & \colhead{Radio counterpart\tablenotemark{l}}  & 
 	    \colhead{S$_{\mathrm{radio}}$\tablenotemark{m}} & \colhead{$q_{22}$\tablenotemark{n}}
	   }
	\startdata
   	J000011.09-433316.4 & 0.0462203 & -43.5545611 & 1.31 & 2.85 & 2.67 & 0.0 & 0.0 & 0.39 & D & BZQ & SUMSSJ000011.2-433317 & 70.9 & -1.41\\
   	J000020.40-322101.2 & 0.0850076 & -32.3503443 & 1.35 & 3.26 & 2.47 & 0.0 & 0.0 & 0.52 & C & BZQ & NVSSJ000020-322059 & 520.9 & -1.94\\
  	J000029.07-163620.2 & 0.1211645 & -16.6056221 & 0.39 & 2.01 & 2.45 & 0.53 & 0.0 & 0.0 & C & BZB & NVSSJ000029-163621 & 89.1 & -1.24\\
   	J000047.05+312028.2 & 0.1960452 & 31.3411703 & 0.82 & 2.48 & 2.4 & 0.32 & 0.0 & 0.0 & D & BZB & NVSSJ000047+312027 & 46.4 & -1.43\\
   	J000056.54-402206.4 & 0.235613 & -40.368453 & 0.71 & 2.69 & 2.6 & 0.28 & 0.0 & 0.0 & D & BZB & SUMSSJ000056.7-402208 & 76.0 & -1.36\\
   	J000101.04+240842.5 & 0.2543718 & 24.1451458 & 1.23 & 3.06 & 2.13 & 0.0 & 0.0 & 0.34 & D & BZQ & NVSSJ000101+240842 & 46.6 & -1.26\\
   	J000105.29-155107.2 & 0.2720486 & -15.8520035 & 1.18 & 3.42 & 2.14 & 0.0 & 0.0 & 0.33 & D & BZQ & NVSSJ000105-155106 & 347.5 & -2.01\\
   	J000108.11-373857.1 & 0.2838199 & -37.6492076 & 1.32 & 2.72 & 2.29 & 0.0 & 0.0 & 0.39 & D & BZQ & SUMSSJ000108.0-373901 & 23.1 & -1.12\\
   	J000118.01-074626.9 & 0.3250683 & -7.7741395 & 0.95 & 2.58 & 2.19 & 0.64 & 0.13 & 0.0 & C & BZB & NVSSJ000118-074626 & 208.4 & -1.23\\
   	J000131.63+165413.8 & 0.3818138 & 16.9038342 & 1.21 & 2.92 & 2.45 & 0.0 & 0.0 & 0.59 & C & BZQ & NVSSJ000131+165416 & 63.1 & -1.08\\
   	J000132.22+135258.4 & 0.3842501 & 13.8829081 & 1.07 & 3.0 & 2.48 & 0.0 & 0.0 & 0.23 & D & BZQ & NVSSJ000132+135258 & 74.2 & -1.59\\
   	J000132.34+240230.3 & 0.3847546 & 24.0417769 & 0.85 & 2.14 & 1.84 & 0.81 & 0.0 & 0.0 & B & BZB & NVSSJ000132+240231 & 359.2 & -1.28\\
   	J000132.74-415525.2 & 0.3864515 & -41.9236926 & 0.68 & 2.17 & 2.09 & 0.48 & 0.0 & 0.0 & D & BZB & SUMSSJ000133.1-415526 & 11.6 & -0.84\\
   	J000132.83+145607.9 & 0.3868025 & 14.9355453 & 0.92 & 2.85 & 2.77 & 0.0 & 0.1 & 0.31 & D & BZQ & NVSSJ000132+145609 & 314.6 & -1.69\\
   	J000137.07+431543.9 & 0.4044828 & 43.2622002 & 0.72 & 2.61 & 2.32 & 0.76 & 0.0 & 0.0 & B & BZB & NVSSJ000137+431544 & 61.8 & -0.9\\
    \enddata
    \tablecomments{(a): \wse\ name; (b): Right Ascension (J2000); (c): Declination (J2000); (d): W1-W2 \wse\ color; 
    (e): W2-W3 \wse\ color; (f): W3-W4 \wse\ color; (g): Score for the BZB region of the {\it locus}; (h): Score for the Mixed region of the {\it locus}; (i):
    Score for the BZQ region of the {\it locus}; (j): Class (see Section~\ref{sec:pc});
    (k): Spectral type (see Section~\ref{sec:pc}); (l): Name of the radio counterpart; 
    (m): Radio flux density [mJy]; (n): Radio-loudness parameter $q_{22}$}
\end{deluxetable*} 

\section{Catalog of \wse\ $\gamma$-ray BL Lac candidates}
\label{sec:wisekde}

\begin{figure}[ht]
    \centering
	\includegraphics[scale=0.5]{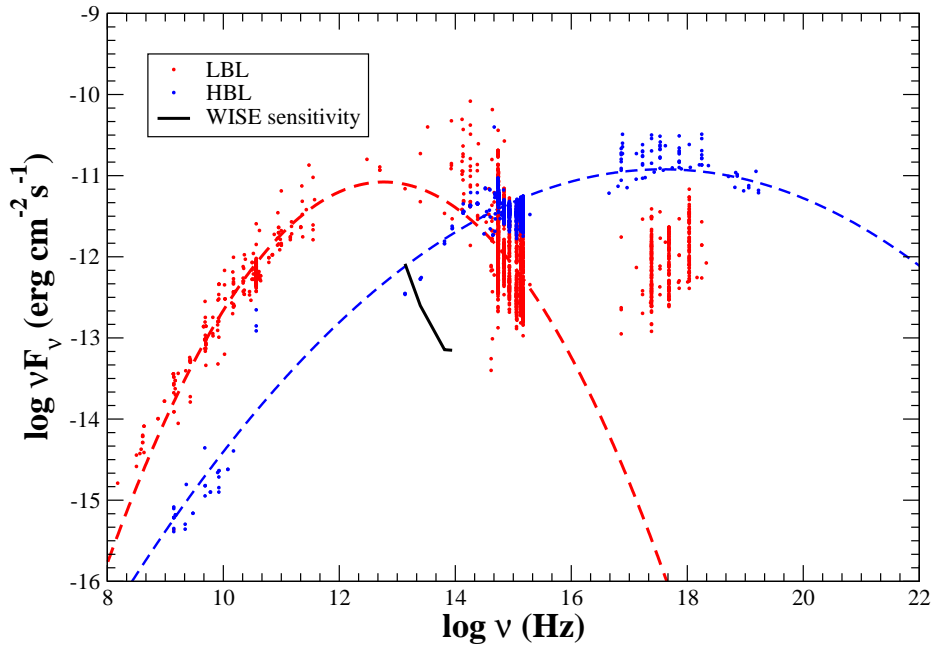}
	\caption{Historical SEDs of the two BZBs 1H 1426+428 (blue points), a TeV HBL, and 
	AO 0235+164 (red points), a typical LBL~\citep[see][respectively]{massaro2008,massaro2008f}. 
	Dashed lines represent the best fit log-parabolic functions used to describe their synchrotron 
	components. The sensitivity limits in the \wse\ 4 bands are shown as a solid black line.} 
	\label{fig:sensitivity}
\end{figure}

As mentioned briefly in Section~\ref{sec:data_wise}, only $\approx 2.5\%$ of the All\wse\ sources are 
detected at $[22]$ $\mu$m. 
For this reason, the WIBRaLS catalog, that requires its members to be detected in all four \wse\ bands, 
will certainly be incomplete for BZBs whose SEDs peak in the X-rays and are not bright 
enough to be detected in the W4 \wse\ band. This occurrence is schematically displayed 
in Figure~\ref{fig:sensitivity}, where the typical shapes of the SEDs of HBLs and LBLs
are plotted together with the \wse\ sensitivity limits for the four \wse\ filters. 

We designed a novel method that relies only on the colors obtained with the W1, W2 and W3 \wse\ bands to 
produce a list of candidate BL Lacs, extracted from the All\wse\ survey. The steps adopted for 
our procedure can be summarized as follows:

\begin{enumerate}
	\item We select all All\wse\ sources detected only in the first three \wse\ bands, having IR colors 
	similar to 
	those of known \fer\ BZBs belonging to the ROMA-BZCat v5.0. 
	\item From this sample, we further select only those sources with a radio counterpart in either one 
	of the NVSS, SUMSS and FIRST radio surveys.
	\item We identify as candidate BZBs sources whose values of the radio-to-infrared 
	flux density ratio are compatible with those of confirmed \fer\ BL Lacs. 
\end{enumerate}

Details on these steps are given in the following sections.

\subsection{\wse\ color selection of BZB-like sources}
\label{sec:kde}

The training set used to identify the \wse\ mid-IR colors of BL Lac objects was built selecting all the \fer\ 
sources belonging to the 3FGL and associated with BZBs in the latest release of the Roma-BZCAT 
catalog~\citep{massaro15a}. We only considered sources whose All\wse\ counterparts are not detected in the
W4 band. The total number of unique sources in the training set is 93. Among these sources, 34 are associated
with a NVSS radio counterpart only, 3 are associated with a FIRST source only and the remaining 56 have both a 
NVSS and FIRST counterpart.

We selected \wse\ sources with IR colors similar to those of our training set by adopting the same procedure 
used in previous analyses~\citep[see e.g.][]{massaro11,paggi13,massaro13a}, based on the Kernel Density Estimation 
(KDE). KDE is a non-parametric procedure that estimates the Probability Density Function (PDF) of a multivariate 
distribution with no assumption on the properties of the parent population. The KDE depends on only one 
parameter, i.e., the bandwidth of the kernel of the density estimator, which is qualitatively analogous to the 
window size for one-dimensional running average.

\begin{figure}[ht] 
    \centering
    \includegraphics[scale=0.5]{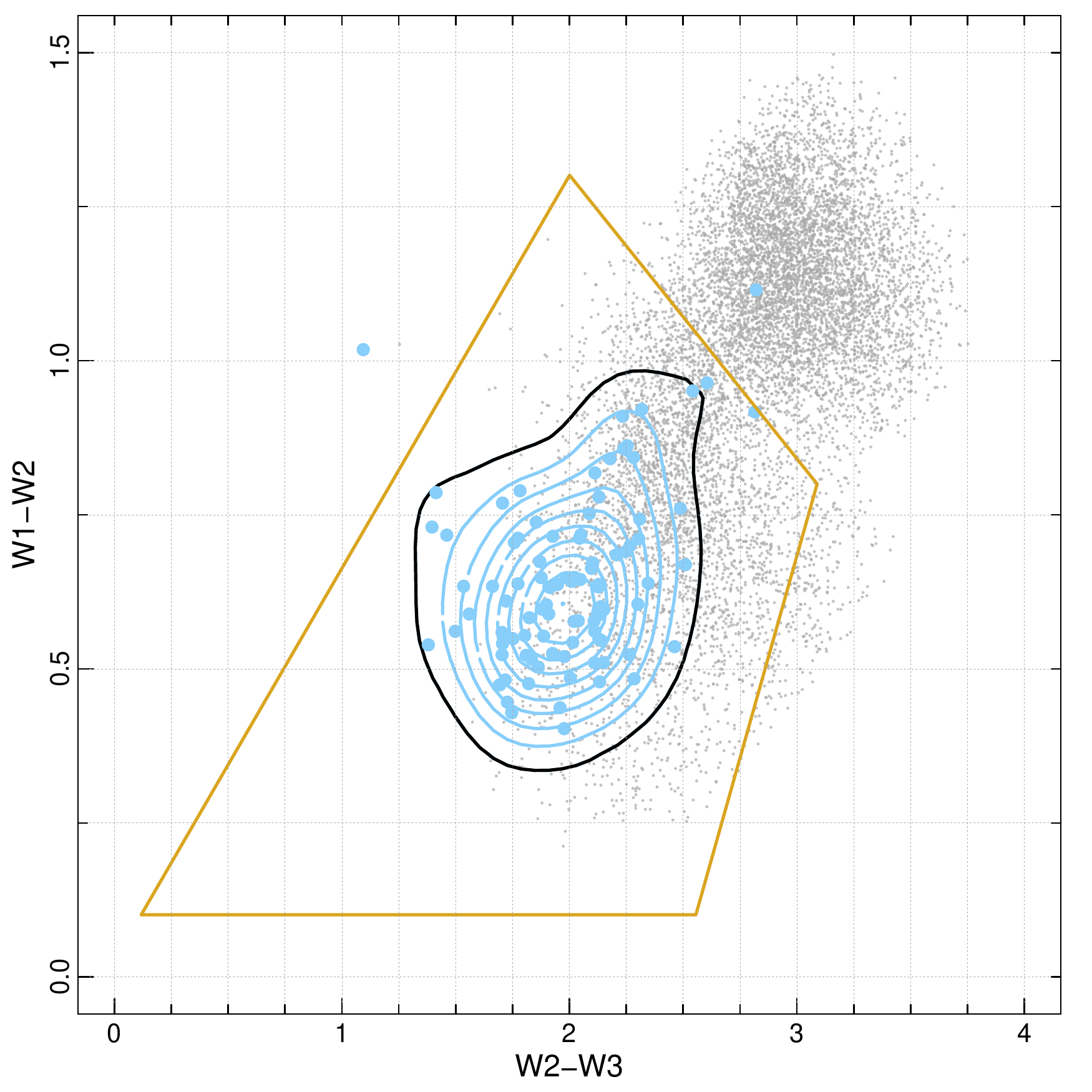}
	\caption{\wse\ W2-W3 vs W1-W2 color-color diagram. The 93 sources in the training set are shown by
	light blue circles, and the black line is the contour containing 90\% of the sources in the training
	set and used to select the candidate KDE BL Lacs. Sources in the WIBRaLS catalog are displayed 
	in the background for comparison (small gray circles). The yellow polygon shows the region in the 
	\wse\ colors plane sources from the 2WHSP catalog are located~\citep{chang2017} 
	(see Section~\ref{subsec:literature}).}
    \label{fig:contours}
\end{figure}

We applied the KDE to the 2-dimensional distribution of training set sources in the \wse\ W2-W3 vs W1-W2 
color-color plane to determine its PDF. Then, we selected infrared sources in the All\wse\ catalog not detected in 
the W4 passband and detected 
in the other three filters, whose colors are located within the isodensity contour enclosing 90\% of the BL Lac 
training set. Only sources whose colors uncertainties ellipses are entirely contained within the 90\% isodensity
contour were retained. Figure~\ref{fig:contours} shows the distribution of the training set in the \wse\ color 
plane, together
with the iso-density contours determined by the KDE method, with the 90\% contour displayed as a thick black line.  
The projections on this color-color diagram of the WIBRaLS2 catalog is also shown for reference.

The total number of \wse\ sources extracted applying this method is $\sim$14406, corresponding to $\sim\!0.01\%$ 
of the parent sample of All\wse\ sources not detected only in W4.

\subsection{Radio counterparts}
\label{sec:radio}

We further select possible BL Lacs candidates by searching for radio counterparts of the sample of sources selected
based on their \wse\ colors. This procedure obviously misses ``radio weak BL Lacs'' but to date they are  
extremely rare and their associations with \fer\ sources has not yet been verified~\citep{massaro17,bruni18}. 

We crossmatched the color-selected \wse\ sources with the radio surveys NVSS, SUMSS and FIRST using 
the same maximum association radii established as described in Section~\ref{sec:radctp}. We 
found a total of 17826 sources with at least one radio counterpart: 5532 are associated with a FIRST 
source within 3\arcsec.4, 
10166 with a NVSS source within 10\arcsec and the remaining 2128 with a SUMSS source with 7\arcsec.4  
(see Table~\ref{tab:KDEBLLacs}). After removing from our list sources with SUMSS or FIRST counterparts 
that are also associated with a NVSS source, there remain 2099 \wse\ sources with a unique FIRST 
counterpart, 1680 \wse\ sources with a unique SUMSS counterpart and the 10166 associated with a NVSS radio 
source, for a total of 13945 sources. Similarly at what was done for the WIBRaLS catalog 
in Section~\ref{sec:radio}, 
in order to avoid possible contamination from radio lobes originating from radio-galaxies, 
we also checked for sources that had from another radio source
closer than 6\arcmin, and, for FIRST radio counterparts only, sources with side lobe 
probability $\geq$ 0.05. These constraints reduced the number of sources to 2049 sources 
with a unique FIRST counterpart, 1671 sources associated to a unique SUMSS source and 10084 w
ith a radio counterpart in NVSS, for a total of 13804 candidates.

\subsection{Infrared-to-radio ratio selection}
\label{sec:ratios}

The last step of the procedure to select \wse\ BL Lac candidates is based on the characterization of the 
distribution of their infrared-to-radio ratios $q_{12}$, similar to the $q_{22}$ parameter used to select 
WIBRaLS sources~\ref{sec:radioloud}. We define the parameter $q_{12}$ as:

\begin{equation}
	q_{12}\!=\!\log{(S_{12\mu\!m}/S_{\mathrm{radio}})}
	\label{eq:rho}
\end{equation} 

\noindent i.e., the logarithm of the ratio between the \wse\ flux density measured in the W3 passband and 
the radio flux density. Flux density at 20 cm has been used to calculate $q_{12}$ of sources with a unique 
counterpart in the NVSS and FIRST surveys. 
Given the flatness of the radio spectrum of BL Lacs~\citep{healey07,massaro13a} and our estimate of the 
$\Delta q_{22}\!\approx\!-0.07$ for the \wse\ -selected 
sources with both SUMSS and NVSS radio counterparts (see Section~\ref{sec:radioloud}), 
we assume a flat spectrum ($\alpha_{\mathrm{radio}}\!=\!0$) and use the flux density at 36 cm to
calculate the $q_{12}$ parameter for sources with SUMSS counterparts.~\footnote{We estimate that for 
radio spectra with 
$\alpha_{\mathrm{radio}}\!=\!\pm0.1$, the effect on the values of 
$q_{12}(843~\mathrm{MHz})$ calculated at 36 cm for SUMSS counterparts 
would be a factor ranging from $\approx\!0.95$ to $\approx\!1.05$. In the case of a 
much larger spectral index $\alpha_{\mathrm{radio}}\!=\!0.5$, the corrective factor would still be 
$\approx\!0.77$.}

We consider BL Lac candidates those sources whose $q_{12}$ values are consistent with the observed $q_{12}$ 
distribution of the training set sources. The $q_{12}$ values for sources with a FIRST or NVSS radio 
counterpart (59 and 90 respectively) have been used 
separately to determine the intervals of acceptable $q_{12}$ values for both surveys, as in both surveys flux 
densities are measured at the same wavelength (20 cm). 

\begin{figure}[ht]
    \centering
 	\includegraphics[scale=0.5]{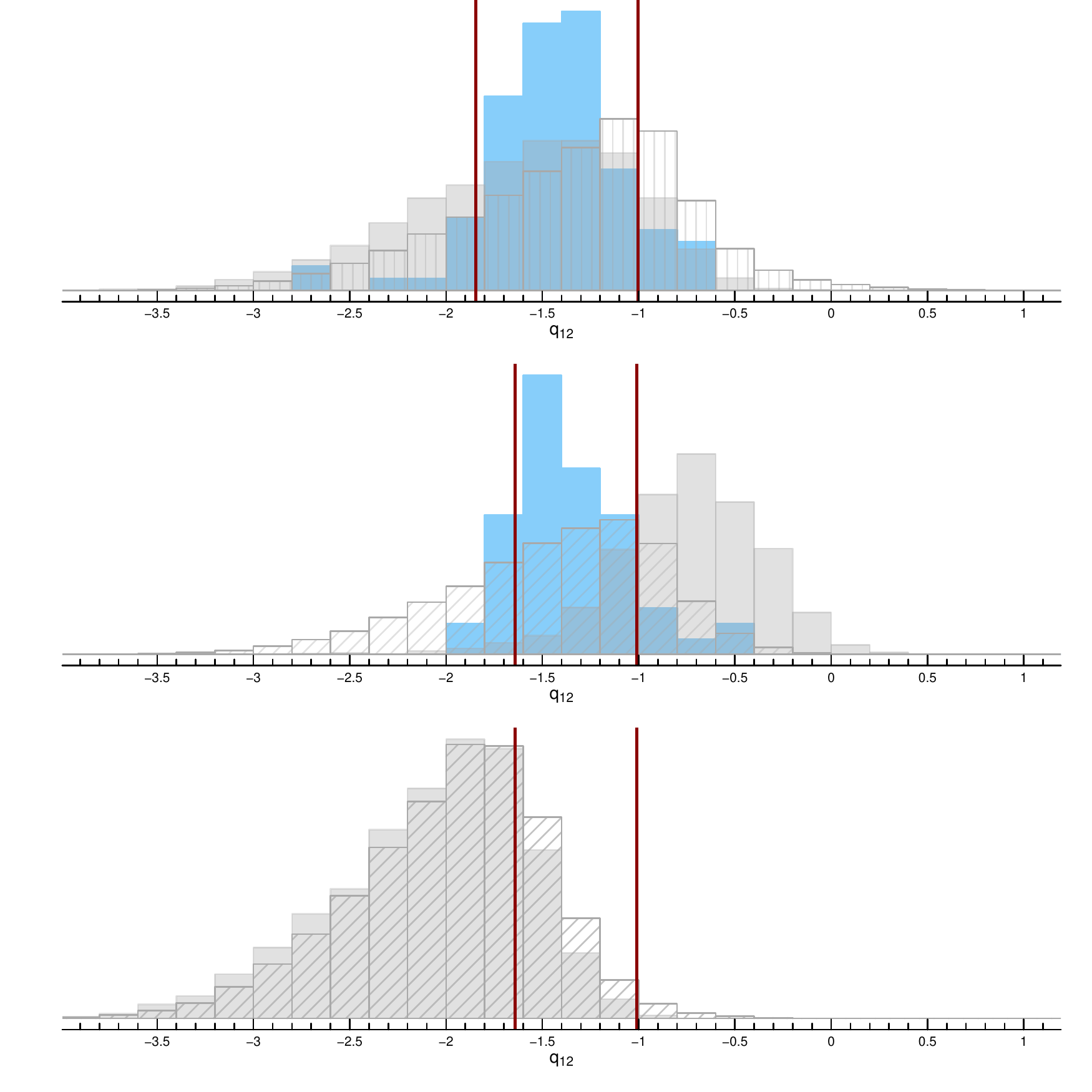}
	\caption{Upper panel: the light blue histogram shows the distributions of $q_{12}$ values for 
	confirmed $\gamma$-ray emitting BL Lacs 
	(training set sources for the KDE candidate BL Lacs) with radio counterpart in the FIRST radio surveys. 
	The solid and dashed gray histograms show the $q_{12}$ distributions for a random sample of \wse\
	sources associated with a FIRST counterpart and for the \wse\ -color selected candidate KDE BL Lacs,
	respectively. Mid panel: same as for upper panel for NVSS radio counterparts. Lower panel: $q_{12}$ 
	distributions for a random sample of \wse\ sources associated with a SUMSS radio source and for 
	\wse\ -color selected candidate KDE BL Lacs (no training set sources with SUMSS radio counterparts
	are available). In all panels,
	vertical red lines display the 10-th and the 90-th percentiles of the $q_{12}$ distribution of 
	the training set used to select the final sample of sources in the KDEBLLACS catalog 
	(see Section~\ref{sec:ratios}).}
    \label{fig:radioIR}
\end{figure}

Figure~\ref{fig:radioIR} shows the distributions of the $q_{12}$ values of the BL Lacs
training set sources (light blue histograms), control samples including random \wse\ 
sources with radio counterparts (solid gray histograms) and of the candidate BL Lacs 
selected according to their \wse\ colors as described in Section~\ref{sec:kde} (dashed gray 
histograms). FIRST (upper panel), NVSS (mid panel) and 
SUMSS (lower panel) radio counterparts are shown separately; the light blue histogram is missing from the
SUMSS panel as no source in the training set has a radio counterpart from the SUMSS survey. It is
also interesting to notice that the peaks of the distributions of $q_{12}$ values for the control sample and 
the \wse\ -selected candidates with SUMSS radio associations (lower panel in Figure~\ref{fig:radioIR})
are significantly shifted towards smaller values of $q_{12}$ relative to the NVSS and FIRST distributions. 
This difference is due to relative shallowness of the SUMSS survey, whose sensitivity allows the detection of
sources with minimum density flux at 36 cm $S_{\mathrm{radio}}^{\mathrm{SUMSS}}(36~\mathrm{cm})\!=\!5$ mJy, larger than the 
$\sim\!2$ mJy and $\sim\!0.2$ mJy for NVSS and FIRST, respectively. 

We defined the lower and upper limits of $q_{12}$ values used to select the candidate BL Lacs as
the 10-th and 90-th percentiles of the $q_{12}$ distribution of the training set (red lines in 
Figure~\ref{fig:radioIR}). The $q_{12}$ intervals used to select the candidates are between -1.85 and -1 for 
NVSS training set sources and between -1.64 and -1.01 for sources with FIRST counterparts.
Given the lack  of training set BL Lacs with SUMSS counterpart, we have conservatively defined 
the upper and lower thresholds for the $q_{12}$ selection of SUMSS BL Lac candidates 
as the highest and lowest values of the lower and upper thresholds 
on the $q_{12}$ distribution of training set sources with a FIRST or 
NVSS counterpart, respectively. As a result, the interval of allowed $q_{12}$ values for SUMSS counterparts 
ranges between -1.64 and -1.
After applying the $q_{12}$ selections to the sample of 13804 \wse-radio selected sources, we obtain 
5941 candidates, split in 327 associated to 
a FIRST source, 5310 with a NVSS counterpart and the remaining 305 associated with a 
SUMSS source. 

Then, as final step, we discarded all sources in the catalog located at Galactic latitudes $|b|\!<\!10\deg$,
since $\gamma$-ray sources at low galactic latitudes suffer from a higher detection threshold due 
to a higher Galactic diffuse emission background~\citep{ackermann15}. This constraint reduces the total number
of sources in the catalog of KDE-selected BL Lac candidates to 5579, due to the exclusion of 363 sources 
with a NVSS radio counterpart. In what follows, the sources in the catalog 
of KDE-selected candidate BL Lacs will be called KDEBLLACS. The break-down of the number of BL Lacs candidates 
selected at different stages of the procedure as a function of the different radio surveys is displayed 
in Table~\ref{tab:KDEBLLacs}, while the properties available for a sample of sources
in the final KDEBLLACS catalog is shown in Table~\ref{tab:KDEBLLacs}

\begin{deluxetable}{lcccc}
	\tablecaption{Break-down of the number of \wse\ sources identified as BL Lacs candidate at each 
	step of the selection described in Section~\ref{sec:wisekde}, split by provenance of the radio 
	counterparts.\label{tab:KDEBLLacs}}
	\tablehead{
	    & \colhead{NVSS}	& \colhead{FIRST}   & \colhead{SUMSS}   & 	 \colhead{Total}	\\
	    }
    \startdata
	\wse\ selection	        &	10166   	& 5532	&2128		& 	17826	\\
	duplicates removal		&	10166	    & 2099	& 1680		&	13945	\\
	close sources removal	&	100084		& 2049	& 1671		&	13804	\\ 	
	$q_{12}$ selection	    &	5310		& 327	& 305		&	5942	\\
	$\|b\|$ selection   	&	4947		& 327	& 305		&	5579	\\
	\enddata
\end{deluxetable} 

\begin{deluxetable*}{lrrrrlrr}
	\tablecaption{Sample of rows of the catalog of KDEBLLACS.\label{tab:cat_kde}}
	\tablehead{
	    \colhead{All\wse\ name\tablenotemark{a}} & \colhead{R.A.\tablenotemark{b}} & \colhead{Dec.\tablenotemark{c}}  & 
	    \colhead{W1-W2\tablenotemark{d}} & \colhead{W2-W3\tablenotemark{e}} & \colhead{Radio name\tablenotemark{f}} & 
	    \colhead{S$_{20cm}$\tablenotemark{g}} & 
	    \colhead{q$_{12}$\tablenotemark{i}}
	    }    
	\startdata
	  J000007.63+420725.5 & 0.0318093 & 42.1237527 & 0.35 & 1.92 & NVSSJ000007+420722 & 18.2 & -1.5\\
	  J000010.29-363405.2 & 0.042887 & -36.5681267 & 0.53 &  2.33 & NVSSJ000010-363407 & 6.2 & -1.3\\
	  J000056.22-082742.0 & 0.2342813 & -8.4616809 & 0.43 &  1.94 & NVSSJ000056-082747 & 14.4 & -1.6\\
	  J000116.37+293534.5 & 0.3182368 & 29.5929424 & 0.61 &  2.46 & NVSSJ000116+293534 & 3.5 & -1.02\\
	  J000126.44+733042.6 & 0.3601711 & 73.5118347 & 0.77 &  2.22 & NVSSJ000126+733042 & 23.6 & -1.63\\
	  J000137.86-103727.3 & 0.4077672 & -10.6242584 & 0.40 & 2.07 & NVSSJ000137-103727 & 10.2 & -1.41\\
	  J000147.28+455015.2 & 0.4470018 & 45.8375759 & 0.78 &  1.91 & NVSSJ000147+455016 & 4.2 & -1.22\\
	  J000236.06-081532.4 & 0.6502775 & -8.2590058 & 0.71 &  2.11 & NVSSJ000236-081533 & 28.3 & -1.61\\
	  J000302.99-105638.1 & 0.7624893 & -10.9439389 & 0.67 & 2.36 & NVSSJ000302-105637 & 18.2 & -1.61\\
	  J000311.94-070144.3 & 0.7997588 & -7.0289838 & 0.61 & 1.98 & FIRST J000311.9-070144 & 4.17 & -1.01\\
	\enddata
    \tablecomments{(a): \wse\ name; (b): Right Ascension (J2000); (c): Declination (J2000); (d): W1-W2 \wse\ color; 
    (e): W2-W3 \wse\ color; (f): Name of the radio counterpart; 
    (h): Radio flux density [Jy]; (n): Radio-loudness parameter q$_{12}$}
\end{deluxetable*} 

\section{Discussion}
\label{sec:catalogs}

The full characterization of the two catalogs presented in this paper would require
optical, spectroscopic follow-up observations to confirm the nature of the candidates, and will be 
discussed in a future paper. In this Section, we examine the global spatial and 
\wse\ photometric properties of the WIBRaLS2 and KDEBLLACS catalogs, and discuss how they compare with the 
most recent catalogs of \fer\ $\gamma$-ray sources. 

\begin{figure*}[ht]
	\includegraphics[width=0.5\linewidth]{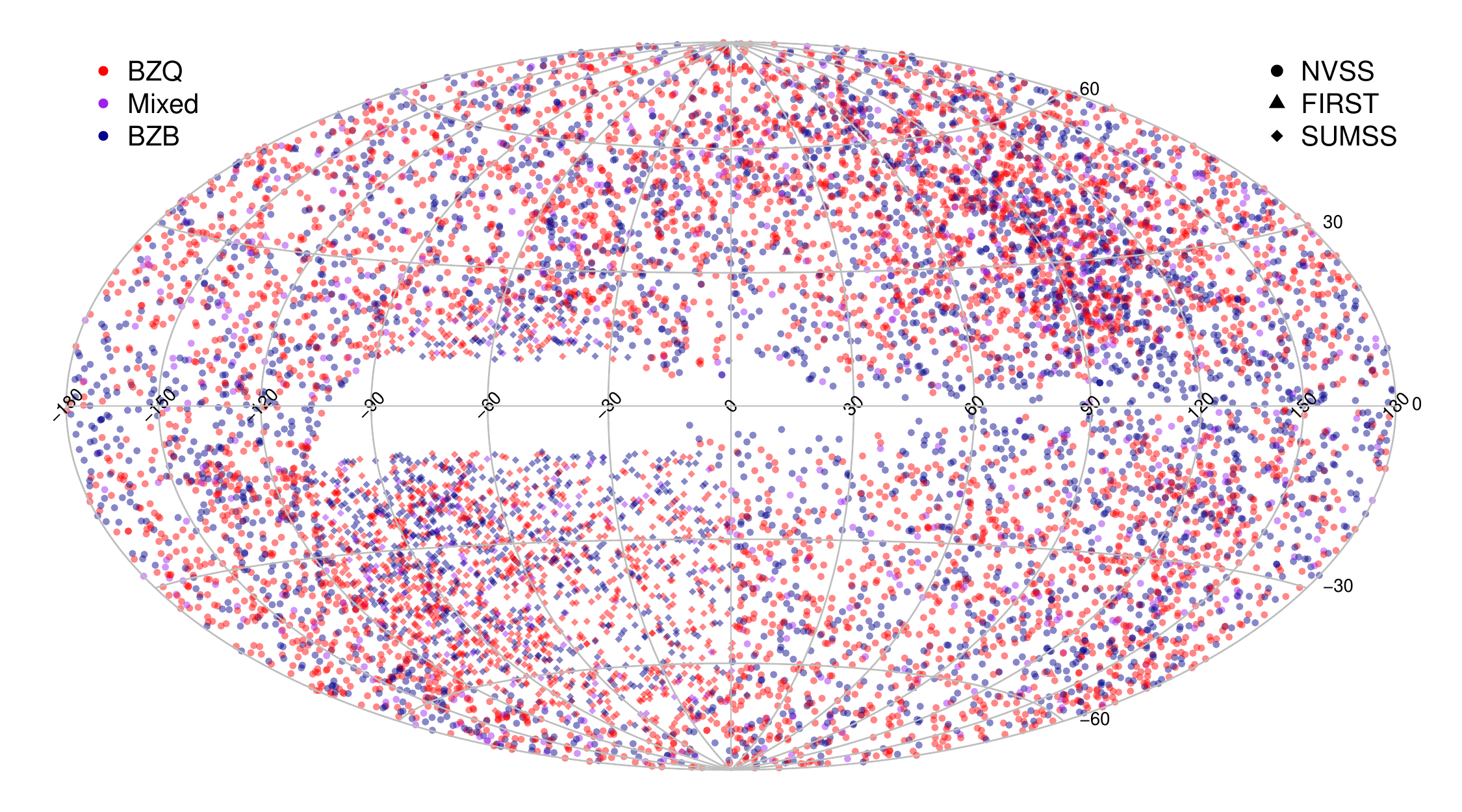}
	\includegraphics[width=0.5\linewidth]{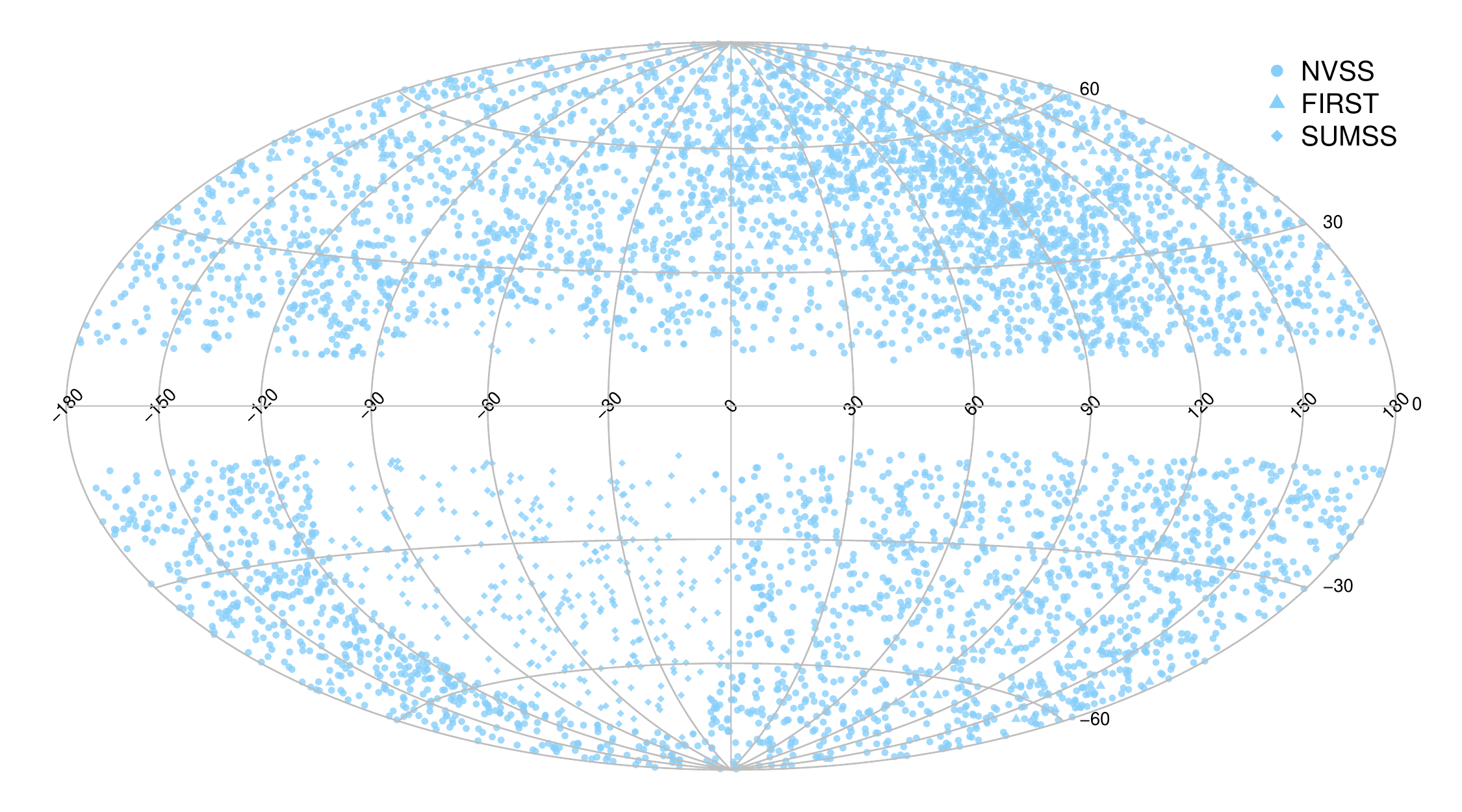}
	\caption{Left panel: Aittof projection of the distribution in galactic coordinates of the 
	WIBRaLS2 sources. The provenance of the radio counterpart and \wse\ spectral class
	of the candidate blazars
	are encoded in the shape and color of the symbols, respectively. Right panel: Aitoff projection
	of the distribution in galactic coordinates of the BL Lacs candidates selected with the KDE
	method. The provenance of the radio counterpart is encoded in the shape of the symbols.}
\label{fig:sky_distribution}
\end{figure*}

The Aitoff projections of the sky positions in galactic coordinates of the WIBRaLS2 and KDEBLLACS catalogs 
are shown in Figure~\ref{fig:sky_distribution}. The coverage of the WIBRaLS2 catalog is mostly uniform 
across the sky (left panel in Figure~\ref{fig:sky_distribution}), with the exception of a region along
the galactic plane where radio sources are not available from any of the three surveys used. Two regions 
of higher density can be observed where the FIRST and SUMSS surveys
overlap with the NVSS coverage, respectively north and south of the Galactic plane. In the right 
panel, the sky distribution of the KDEBLLACS catalog features prominently a lower density of sources
associated to SUMSS radio counterparts and the lack of sources due to the Galactic latitude selection 
described in Section~\ref{sec:ratios}. 

\begin{figure*}[ht]
	\includegraphics[width=\linewidth]{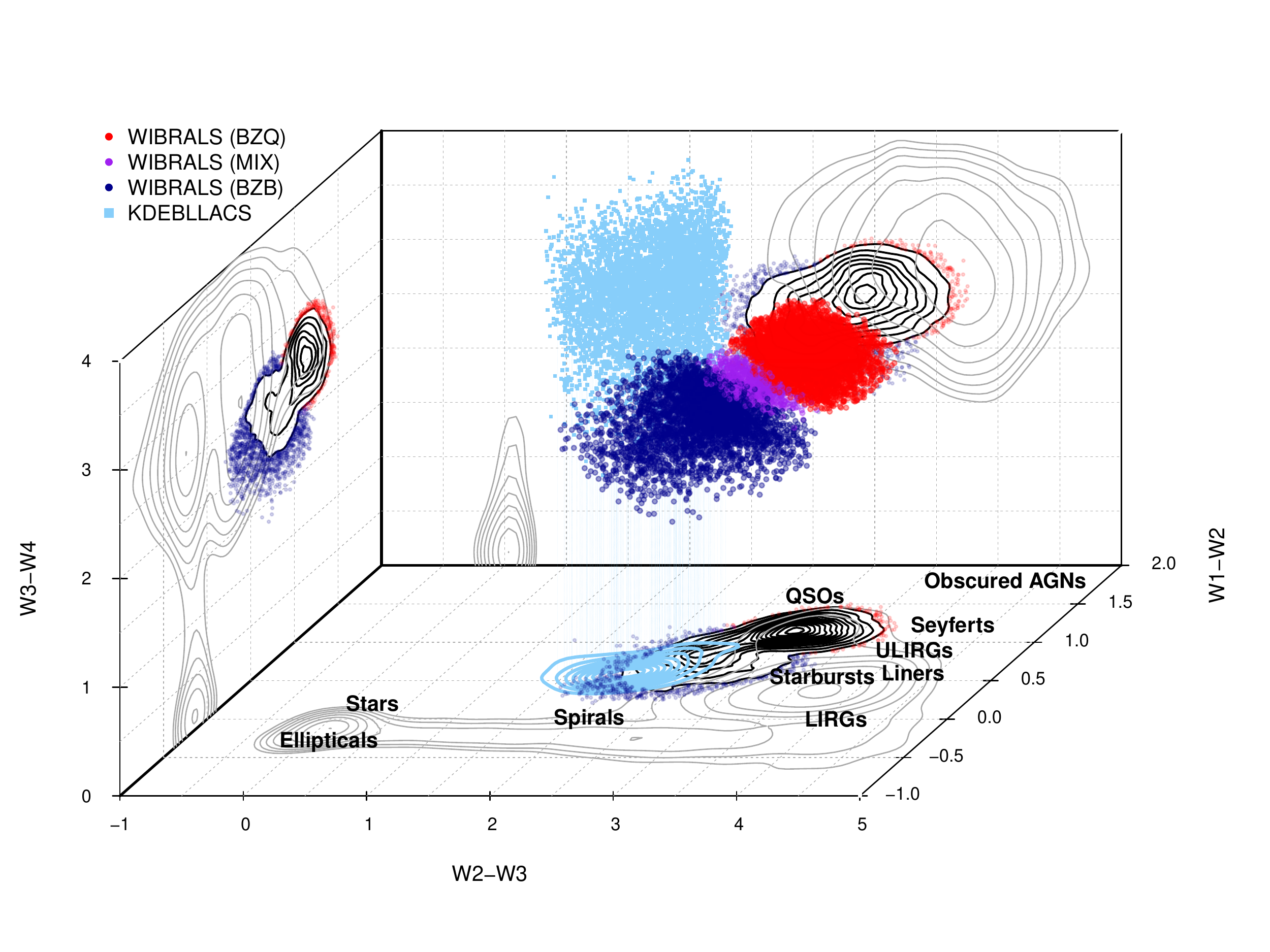}
	\caption{Distribution of the WIBRaLS2 and KDEBLLACS (light blue) catalogs in the three-dimensional \wse\ color space. 
	WIBRaLS2 sources are color-coded 
	according to their \wse\ spectral classification, while the three-dimensional positions of the 
	KDEBLLACS sources (light blue), given their non-detection 
	in the W4 filter, is visualized by using the lower limit in the W4 magnitude and by segments originating
	from these points and delimiting the 3-D volume in the color space where these sources may be actually located. 
	The isodensity contours of the distributions of WIBRaLS2 (black) and 
	KDEBLLACS (blue light) samples are shown in the W2-W3 vs W1-W2 plane, while the gray 
	lines on the three color-color planes represent the projected isodensity contours associated 
	with 10 log-spaced levels of a sample of \wse\ random sources (including both sources detected and not
	detected at 22 $\mu$m. The approximate locations of 
	different classes of sources in the W2-W3 vs W1-W2 color-color plane, according to~\citet{wright10},
	are also shown for guidance.}
    \label{fig:3dcatalogs}
\end{figure*}

It is interesting to compare the regions of the \wse\ color space occupied by the WIBRaLS2 and 
KDEBLLACS catalogs. The location of the sources belonging to the above catalogs in the \wse\ 
color space is displayed in Figure~\ref{fig:3dcatalogs}.
The two samples occupy partially overlapping but distinct regions of the three
dimensional \wse\ color space. Since KDEBLLACS are not detected in the W4 passband by definition, 
their positions along the W3-W4 axis cannot be established, but the three-dimensional volume of the 
space potentially occupied by these sources can be determined by using the upper limits on their W4 
brightness (i.e., lower limits on their W4 magnitude) available in the All\wse\ catalog.

\begin{figure}[ht]
    \centering
	\includegraphics[scale=0.3]{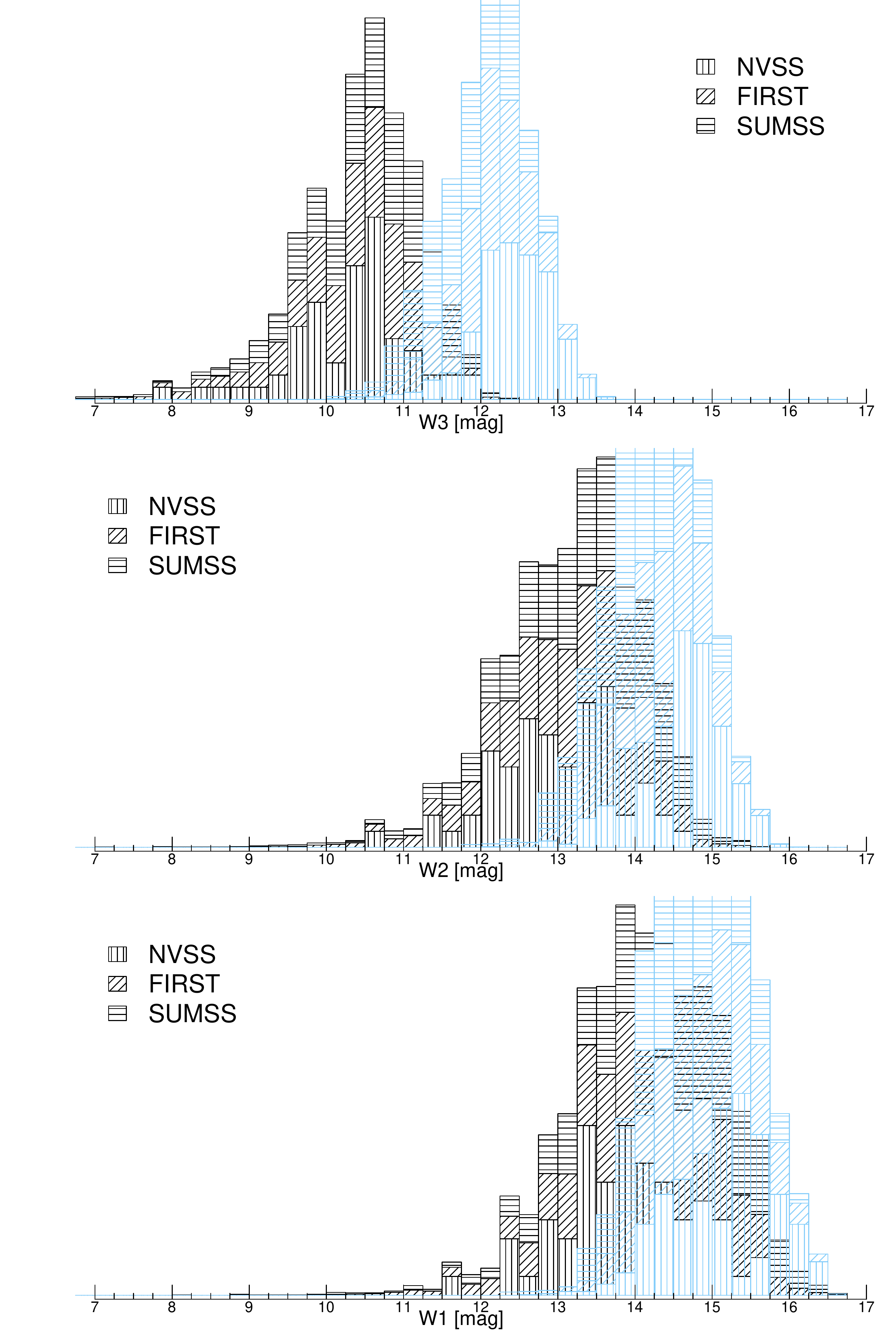}
	\caption{Distribution of magnitudes in the W1 (lower panel), W2 (mid panel) and W3 (upper panel) 
	\wse\ filters for WIBRaLS2 (black histograms) and
	KDEBLLACS (light blue histogram) candidate blazars, broken down by provenance of the 
	radio counterpart.}
\label{fig:histomags}
\end{figure}

The two catalogs of blazar candidates described in this paper are complementary because their members
can differ in spectral properties (see discussion in Section~\ref{sec:wisekde}) and in brightness. 
Figure~\ref{fig:histomags}, that displays the histograms of the magnitude values for the three \wse\
filters W1, W2 and W3 for the WIBRaLS2 and KDEBLLACS samples together, confirms that the KDE-based method 
selects sources that are increasingly fainter than those in the WIBRaLS catalog as the wavelength increase 
going from filter W1 to W3. 

\subsection{Comparison with literature}
\label{subsec:literature}

One of the main goals of the production of catalogs of candidate blazars, like the two discussed
in this paper, is the discovery of multi-wavelength counterparts of unassociated $\gamma$-ray 
sources (UGSs) observed by {\it Fermi}. While the thorough analysis that is necessary to reliably 
associate low energy candidate blazars to UGS~\citep[see, for example][]{dabrusco13,massaro15b,schinzel15}
is beyond the goals of this paper, as a consistency check and in order to assess the potential of the 
two catalogs here discussed to improve the characterization of the currently known $\gamma$-ray sources, we 
compared the WIBRaLS2 and KDEBLLACS catalogs with the most recent catalogs of sources detected by \fer\ LAT. 

We spatially crossmatched the WIBRaLS2 catalog with the 3FGL catalog~\citep{acero15} using the 98\% 
elliptical uncertainty regions for 3FGL sources and a fixed positional uncertainty of 1\arcsec on the 
WIBRaLS2 radio coordinates. 
We found a total of 1049 matches, including the 666 sources in the {\it locus} sample selected 
as WIBRaLS2 members (see Section~\ref{sec:radioloud}). Among the remaining
373 sources, 49 are unassociated; the 320 sources that are associated or identified with a known 
multi-wavelength counterpart in 3FGL are all classified, according to the CLASS1 parameter in 
3FGL~\citep{acero15}, as BL Lacs, FSRQs or BZU (Blazar candidate of Uncertain type), except for 
3 rdg (radio-galaxies) and 3 ssrq (soft-spectrum radio quasars). The crossmatch between the 
KDEBLLACS catalog and 3FGL, using the same positional uncertainties, returned 186 matches (all 
distinct from the WIBRaLS2 crossmatches), with 57 unassociated sources. $\sim95\%$ of the remaining 131
3FGL sources crossmatched with a KDEBLLACS source that are associated or identified, are classified as 
BZU or BL Lacs according to the 3FGL CLASS parameter. The total number of unique crossmatched sources from 
either the WIBRaLS2 or KDEBLLACS catalogs increases to 1757 (1404 and 353, respectively) when 
the preliminary LAT 8-year Point Source List 
(FL8Y)\footnote{https://fermi.gsfc.nasa.gov/ssc/data/access/lat/fl8y/} (which contains a total of 5523 
$\gamma$-ray sources) is used, for a total of 152 unassociated sources. Out of the 1605
associated or identified sources, $\sim 99\%$ are classified as either BL Lacs, FSRQs or BZU.  

Following the same approach described above, we also crossmatched the WIBRaLS2 and KDEBLLACS 
catalogs with the Third Catalog of Hard \fer-LAT Sources~\citep[3FHL;][]{ajello17a}. The spatial
crossmatch returns a total of 807 distinct matches, split in 647 from WIBRaLS2 and 160 from 
KDEBLLACS, with 33 unassociated 3FHL sources and the remaining 774 composed by BL Lacs 
(520, $\sim 68\%$ of the total), 
and blazars of uncertain type (BZU, $\sim 16\%$ of the total). The $83\%$ of the WIBRaLS2 members 
crossmatched with 3FHL sources are classified as candidate BZBs or MIXED candidates, based 
on their \wse\ colors. 

It is also useful to compare the two catalogs of candidate blazars presented in this paper 
with the largest catalog of candidate HSPs available in the literature to date, namely the second 
WISE High Synchrotron Peaked blazar (2WHSP) catalog~\citep{chang2017}, with 1691 entries. The 2WHSP is an 
expansion of the 1WHSP catalog~\citep{arsioli15}, 
and contains HSPs candidates drawn from the All\wse\ catalog that can be associated to radio and X-ray counterparts. 
The 2WSHPs candidates are further selected by requiring that their radio-to-IR and IR-to-X-ray broad-band 
spectral slopes are consistent with those of known, confirmed HSPs sources~\citep{chang2017}, and that the peak frequency 
of the synchrotron emission component of their SEDs $\nu_{\mathrm{peak}}$ is $\!>\!10^{15}$ Hz~\citep{chang2017}. The 
2WHSP contains 1691 unique candidates or confirmed HSPs: 460 of these sources are associated to All\wse\ sources
detected in four bands and 717 sources are associated to sources detected in the first three (W1, W2 and W3)
filters. The remaining sources cannot be compared to the catalogs discussed in this paper because not detected 
in both the W3 and W4 passbands. We determined that 248 of the 460 2WSHPs sources detected in all All\wse\ filters are
also in the new WIBRaLS catalog, while 267 of the 717 sources not detected in W3 have been selected as 
KDEBLLACS. The main cause of difference between our catalogs and the 2WHSP is 
the maximum spatial radius used to crossmatch All\wse\ sources with radio 
counterparts from one of the three radio surveys FIRST, NVSS and SUMSS.~\cite{chang2017} report that 
radio counterparts were selected within 0.3\arcmin\ and 0.1\arcmin\ for NVSS/SUMSS and FIRST catalogs, 
respectively. These radii are significantly larger than the radii used in this catalog, discussed
in Sections~\ref{sec:radctp} and~\ref{sec:radio} (10\arcsec.4, 3\arcsec.4 and 7\arcsec.4 for NVSS, 
FIRST and SUMSS respectively). As a consequence, only 311 All\wse\ sources with a unique radio counterpart 
contained in the sample used to extract the KDEBLLACS can also be found in the 
2WSHP catalog. Another possible source of difference between the KDEBLLACS and the 2WHSP catalogs is 
the extent of the region in the \wse\ color-color diagram where 2WHSP sources are located, which is 
significantly larger than the area where the BL Lacs candidates are selected from 
(see Figure~\ref{fig:contours}).

\subsection{Comparison with the WIBRaLS1 catalog}

The second release of the WIBRaLS catalog, WIBRaLS2, described in this paper, contains 5025 candidate blazars also 
found in the first WIBRaLS (WIBRALS1 hereinafter) catalog~(Paper I). Of the total 7885 members of 
WIBRALS1, 2830 sources are not included in WIBRaLS2 because their $q_{22}$ values (for $\sim 95\%$ of them) 
are larger than the new, more stringent thresholds adopted for BZQs in WIBRaLS2 (see Section~\ref{sec:radioloud}). 
The increase in the size of the training set of confirmed $\gamma$-ray blazars used to define the WIBRaLS 
{\it locus} has led to an increase of the volume in \wse\ color space that has produced 4711 sources in 
the WIBRaLS2 not included in WIBRaLS1.

\section{Summary and conclusions}
\label{sec:summary}

In this paper, we present two new catalogs of blazar candidates, selected on 
the basis of their \wse\ mid-IR colors, their association to a radio counterpart and their 
radio-loudness. These new catalogs contain a combined total of 15196 candidate blazars, 
including both candidate FSRQs and BL Lacs, distributed on $\sim 90\%$ of the sky. 

The second, enhanced release of the WIBRaLS catalog supersedes the original WIBRaLS1~(Paper I) catalog, 
which has been extensively employed to associate \fer\ unidentified sources with their low energy 
counterparts. WIBRaLS2 contains candidate blazars drawn from All\wse\ sources detected
in all four \wse\ passbands with colors similar to those of spectroscopically confirmed, $\gamma$-ray 
emitting blazars, that are associated to radio counterparts and identified as radio-loud. 
Spectral classification as candidate BZBs, BZQs or Mixed candidate blazars, following the ROMA-BZCat~\citep{massaro15a} terminology, 
and derived on the \wse\ color properties is also provided. WIBRaLS2 contains 9541 candidate blazars, 
a $\sim\!25\%$ increase over the first version of WIBRaLS.

The KDEBLLACS catalog complements the WIBRaLS2 by identifying BL Lacs candidates 
that, because of their typical low infrared-to-radio ratios and/or \wse\ brightness (cp. HBLs),  
have gone undetected in the 
W4 \wse\ filter and, as a consequence, cannot be considered for selection in WIBRaLS2. The KDEBLLACS members 
are required to be detected in the three \wse\ filters W1, W2 and W3 and are selected based on their positions
in the W2-W3 vs W1-W2 \wse\ color-color diagrams using the KDE technique. They are also associated to a radio 
counterpart and identified as radio-loud according to the $q_{12}$ radio-to-MIR spectral parameter. We select
5579 sources in the KDEBLLACS. 

Previous samples of candidate blazars selected on the basis of the IR colors, eventually combined with 
radio and/or multifrequency observations~\citep[see e.g.][]{massaro13d,arsioli15,maselli15,massaro16} have been 
used by the \fer-LAT collaboration for 
the preparation of the 3FGL, the Third Catalog of Active Galactic Nuclei~\citep[3LAC;][]{ackermann15} and the 
Second and the Third Catalog of Hard \fer-LAT Sources~\citep[2FHL and 3FHL;][respectively]{ackermann16,ajello17a}.

This work will contribute to a more comprehensive understanding of the diverse and fascinating $\gamma$-ray 
sky as observed by {\it Fermi}. In particular, the community will benefit from the WIBRaLS2 and KDEBLLACS 
catalogs of candidate blazars presented 
in this paper and the subsequent programs of follow-up spectroscopic observations needed to confirm their nature and, 
possibly, determine their redshifts by using them to: (i) improve our knowledge of the luminosity function of BL 
Lacs~\citep[see e.g.,][]{ajello14}; (ii) select potential targets for the Cherenkov Telescope Array (CTA) 
as shown by~\citet{massaro13e,arsioli15} ; (iii) obtain more stringent 
limits on the dark matter annihilation in sub-halos~\citep[see e.g.,][]{zechlin12,berlin14}; (iv) search for 
counterparts of new flaring $\gamma$-ray sources~\citep[see e.g.,][]{bernieri13} and of high energy 
neutrino emission~\citep[see e.g.,][]{icecube2018}; (v) test new $\gamma$-ray 
detection algorithms~\citep[see e.g.,][]{campana15,campana16,campana17}; (vi) and, finally, perform 
population studies of the remaining UGSs~\citep[see e.g.,][]{acero13}.

\noindent R.D'A. is supported by NASA contract NAS8-03060 (Chandra X-ray Center).

The work of F.M. and A.P. is partially supported by the ``Departments of 
Excellence 2018-2022'' Grant awarded by the Italian Ministry of Education, 
University and Research (MIUR) (L. 232/2016) and made use of resources
provided by the Compagnia di San Paolo for the grant awarded on the BLENV
project (S1618\_L1\_MASF\_01) and by the Ministry of Education, 
Universities and Research for the grant MASF\_FFABR\_17\_01. F.M. 
also acknowledges financial contribution from the agreement ASI-INAF 
n.2017-14-H.0 while A.P. the financial support from the Consorzio 
Interuniversitario per la fisica Spaziale (CIS) under the agreement 
related to the grant MASF\_CONTR\_FIN\_18\_02.

F.R. acknowledges support from FONDECYT Postdoctorado 3180506 and CONICYT project 
Basal AFB-170002.

V.C. is partially supported by the CONACyT research grant 280789

%% grants
%H. A. Smith acknowledges partial support from NASA/JPL grant RSA 1369566.
%The work by G. Tosti is supported by the ASI/INAF contract I/005/12/0.
% HEASARC

This research has made use of data obtained from the high-energy Astrophysics Science Archive Research 
Center (HEASARC) provided by NASA's Goddard Space Flight Center.
% SIMBAD and NED
The SIMBAD database operated at CDS, Strasbourg, France; the NASA/IPAC Extragalactic Database
(NED) operated by the Jet Propulsion Laboratory, California Institute of Technology, under contract 
with the National Aeronautics and Space Administration.
% NVSS
Part of this work is based on the NVSS (NRAO VLA Sky Survey): The National Radio Astronomy Observatory is 
operated by Associated Universities, Inc., under contract with the National Science Foundation and on the VLA 
low-frequency Sky Survey (VLSS).
% SUMSS
The Molonglo Observatory site manager, Duncan Campbell-Wilson, and the staff, Jeff Webb, Michael White and 
John Barry, are responsible for the smooth operation of Molonglo Observatory Synthesis Telescope (MOST) and the 
day-to-day observing programme of SUMSS. The SUMSS survey is dedicated to Michael Large whose expertise and 
vision made the project possible. The MOST is operated by the School of Physics with the support of the Australian 
Research Council and the Science Foundation for Physics within the University of Sydney.
% \wse\
This publication makes use of data products from the Wide-field Infrared Survey Explorer, 
which is a joint project of the University of California, Los Angeles, and 
the Jet Propulsion Laboratory/California Institute of Technology, 
funded by the National Aeronautics and Space Administration.
% 2MASS
This publication makes use of data products from the Two Micron All Sky Survey, which is a joint project of the University 
of Massachusetts and the Infrared Processing and Analysis Center/California Institute of Technology, funded by the 
National Aeronautics and Space Administration and the National Science Foundation.
% TOPCAT
TOPCAT\footnote{\underline{http://www.star.bris.ac.uk/$\sim$mbt/topcat/}} 
\citep{taylor05} for the preparation and manipulation of the tabular data and the images.

\end{document}